\documentclass{elsart}

\usepackage{graphicx}
\usepackage{pxfonts}
\usepackage{lineno}

\usepackage{amssymb}
\usepackage{footnote}
\usepackage{multirow}
\usepackage{varwidth}
\usepackage{array}
\usepackage{epstopdf}
\usepackage{url}
\usepackage{color} 

\journal{}

\begin{document}

\thispagestyle{empty}
\begin{Large}
\textbf{DEUTSCHES ELEKTRONEN-SYNCHROTRON}

\textbf{\large{Ein Forschungszentrum der Helmholtz-Gemeinschaft}\\}
\end{Large}

DESY 17-143

September 2017

\begin{eqnarray}
\nonumber &&\cr \nonumber && \cr \nonumber &&\cr
\end{eqnarray}
\begin{eqnarray}
\nonumber
\end{eqnarray}
\begin{center}
\begin{Large}
\textbf{Relativity and Accelerator Engineering}
\end{Large}
\begin{eqnarray}
\nonumber &&\cr \nonumber && \cr
\end{eqnarray}

\begin{large}
Gianluca Geloni,
\end{large}
\textsl{\\European XFEL GmbH, Schenefeld}
\begin{large}

Vitali Kocharyan and Evgeni Saldin
\end{large}
\textsl{\\Deutsches Elektronen-Synchrotron DESY, Hamburg}
\begin{eqnarray}
\nonumber
\end{eqnarray}
\begin{eqnarray}
\nonumber
\end{eqnarray}
ISSN 0418-9833
\begin{eqnarray}
\nonumber
\end{eqnarray}
\begin{large}
\textbf{NOTKESTRASSE 85 - 22607 HAMBURG}
\end{large}
\end{center}
\clearpage
\newpage

\begin{frontmatter}



\title{Relativity and Accelerator Engineering}

\author[XFEL]{Gianluca Geloni,}
\author[DESY]{Vitali Kocharyan,}
\author[DESY]{Evgeni Saldin}
\address[XFEL]{European XFEL GmbH, Schenefeld, Germany}
\address[DESY]{Deutsches Elektronen-Synchrotron (DESY), Hamburg, Germany}

\begin{abstract}
From a geometrical viewpoint, according to the theory of relativity, space and time constitute a four-dimensional continuum with pseudo-Euclidean structure. This has recently begun to be a practically important statement in accelerator physics. An X-ray Free Electron Laser (XFEL) is in fact the best, exciting example of an engineering system where improvements in accelerator technology makes it possible to develop ultrarelativistic macroscopic objects with an internal fine structure, and the theory of relativity plays an essential role in their description. An ultrarelativistic electron bunch modulated at nanometer-scale in XFELs has indeed a macroscopic finite-size of order of 10 $\mu$m. Its internal, collective structure is characterized in terms of a wave number vector. Here we will show that a four-dimensional geometrical approach, unusual in accelerator physics, is needed to solve problems involving the emission of radiation from an ultrarelativistic modulated electron beam accelerating along a curved trajectory. We will see that relativistic kinematics enters XFEL physics in a most fundamental way through the so-called Wigner rotation of the modulation wave number vector, which is closely associated to the relativity of simultaneity. If not taken into account, relativistic kinematics effects would lead to a strong qualitative disagreement between theory and experiments. In this paper, several examples of relativistic kinematics effects, which are important for current and future XFEL operation, are studied. The theory of relativity is applied by providing details of the clock synchronization procedure within the laboratory frame. This approach, exploited here but unusual in literature, is rather "practical", and should be acceptable to accelerator physicists.
\end{abstract}

%
%
%
\end{frontmatter}


\section{ Introduction }

The primary importance of the geometrical view of the space-time in the theory of special relativity is well-known, and is underlined in several modern treatments, see e.g. \cite{RM,GF,PR,SR,RCM,CT}. The fact that space and time form a four-dimensional continuum with pseudo-Euclidean geometry constitutes the essence of the theory of special relativity, and has consequences on all physical phenomena. Once accepted as a postulate, the principle of relativity follows as its consequence, at variance with more widespread formulations where the geometrical structure of space-time is derived as a consequence of the principle of relativity. In particular, dynamics and electrodynamics equations can be expressed as tensor equations in Minkowski space-time, and automatically include the principle of relativity.

An X-ray free electron laser (XFEL) is currently the best, exciting example of an engineering system in which the description of space-time in terms of 4D geometry plays an essential role. In XFELs, recent improvements in accelerator technology make it possible to develop ultrarelativistic macroscopic object with an internal structure. The need for the 4D geometrical approach, unusual in accelerator physics, may be explained by referring to  problems concerning  relativistic particles accelerating along curved trajectories. A correct solution of these problems requires indeed the use of the 4D geometrical approach. In fact, relativistic kinematics enters the physics of ultrarelativistic objects with internal structure, and therefore XFEL physics, in a most fundamental way through the so-called Wigner rotation. As is known, the composition of non-collinear Lorentz boosts does not results in another  boost, but in a Lorentz transformation involving a boost and a spatial rotation, the Wigner rotation \cite{WI,WI1,WI2}. When an ultrarelativistic electron beam modulated in density is accelerated along a curved trajectory, its evolution is treated according to tensor equations in Minkowski space-time, and covariant equations automatically include the Wigner rotation of the modulation wavefront. Because of this, relativistic kinematics shows a surprising effect:  when the electron beam direction changes after a transverse kick, the orientation of the modulation wavefront is readjusted along the new direction of the electron beam. In other words, when the evolution of the electron bunch modulation is treated according to relativistic kinematics, the orientation of the modulation wavefront in the  ultra-relativistic asymptotic is always perpendicular to the electron beam velocity.

We now turn to misunderstandings that, in our view, have been arising over the last half century, i.e. during all the period when particle accelerators have been in operation. It is generally accepted that in order to describe the dynamics of relativistic electrons in a single inertial  frame there is no need to use the laws of relativistic kinematics. It is sufficient to take into account the relativistic dependence of the particles momenta on the velocity. A theoretical treatment of relativistic particle dynamics in the laboratory (lab) reference frame only involves a correction to Newton's second law. Moreover, the solution of the dynamics problem in the lab frame makes no reference to Lorentz transformations. This means that, for instance, within the lab frame the motion of particles looks precisely  the same as predicted  by Newtonian kinematics: relativistic effects do not have a place in this description. In particular, within the lab frame, the motion of particles along curved trajectories follows the corrected Newton equations, and there no Wigner rotation phenomena arise.

This point was expressed by Friedman \cite{Frid}: "Within any single inertial frame, things looks precisely the same as in Newtonian kinematics: there is an enduring Euclidean three-space, a global (i.e. absolute) time $t$, and inertial law of motion. But different inertial frames are related to one another in a non-Newtonian fashion." Similar explanations can be found in various other advanced textbooks. To quote e.g. Ferrarese and Bini \cite{RCM}: "... within a single Galilean\footnote{With the term 'Galilean', the authors of \cite{RCM} mean 'inertial'} frame, the time is an absolute quantity in special relativity also. As a consequence, \textit{if no more than one frame is involved}, one would not expect differences between classical and relativistic kinematics. But in the relativistic context there are differences in the transformation laws of the various relative quantities (of kinematics or dynamics), when passing from one reference frame to another."

This means that in the lab frame, relativistic kinematics effects like the Wigner rotation are not included into conventional particle tracking theory, used both in accelerator and plasma physics. One can explain this fact by noting that the dynamical evolution in the lab frame is based on the use of the lab frame time $t$ as a independent variable, independent in the sense that $t$ is not related to the spatial variables. In conventional particle tracking time differs from space, and a particle trajectory  $\vec{x}(t)$ can be seen from the lab frame as the result of successive Galileo boosts that track the motion of the accelerated particle. The usual Galileo rule for addition of velocities is used to determine the Galileo boosts tracking a particular particle, instant after instant, along its motion.

The use of Galilean transformations within the theory of relativity requires some discussion. Since the inception of special relativity, most researchers assume that the fact that any two inertial frames are related by a Lorentz transformation directly follows from the postulates of the theory of relativity. However, these postulates alone are not sufficient to obtain a Lorentz transformation: in fact, at each point of the two frames, one assumes to have clocks to measure the time and one therefore needs a procedure to synchronize all the clocks in the first frame, and all the clocks in the second one. This procedure is not fixed by the postulates of the theory of relativity.  If synchronization is done by using the Einstein's synchronization convention, which is almost always understood in textbooks, then a Lorentz transformation follow. However, if the same clocks are synchronized following a different synchronization convention, other transformations follow. In particular, a Galilean transformation can be resulting as well. In the case when a Lorentz transformation is selected the metric tensor in the two frames assumes the usual diagonal form, while in the case of a Galilean transformation the metric tensor in the two frames is not diagonal anymore. However, the frames are and remain inertial, as the 4D interval $ds^2$ between events is kept invariant. We must underline, in fact, that here we are only dealing with a choice of coordinates for two inertial frames, and with the mathematical transformations between the two: we are not touching the structure of space-time, which is and remains pseudo-euclidian four-dimensional continuum.

In order to obtain relativistic kinematics effects, and in contrast to conventional particle tracking, one actually needs to solve the dynamics equation in manifestly covariant form by using the coordinate-independent proper time $\tau$ to parameterize the particle world-line in space-time. Relying on the geometric structure of Minkowski space-time, one defines the class of inertial frames and adopts a Lorentz frame with orthonormal basis vectors. Within the chosen Lorentz frame, Einstein's synchronization of distant clocks and Cartesian space coordinates are enforced. Hence, as just remarked, in Lorentz coordinates we have the well-known diagonal Minkowski metric tensor $g_{\mu\nu} = \mathrm{diag}(1, -1, -1, -1)$, and any two Lorentz frames are related by a Lorentz transformation, which preserves the metric tensor components. In the Lorentz lab frame (i.e. the lab frame with Lorentz coordinate system) one thus has a coordinate representation of a particle world-line as ($t(\tau), x_1(\tau), x_2(\tau), x_3(\tau)$). These four quantities basically are, at any $\tau$, components of a four-vector describing an event in space-time. Therefore, if one  chooses the lab time $t$ as a parameter for the trajectory curve, after inverting the relation $t = t(\tau)$, one obtains that the space position vector  of a particle in the Lorentz lab frame has the functional form $\vec{x}_{cov}(t)$.

In the previous discussion we sketched the derivation of a conventional (or 'non-covariant') particle trajectory $\vec{x}(t)$, calculated by solving corrected  Newton's equations, and a covariant particle trajectory $\vec{x}_{cov}(t)$, calculated by projecting world line onto the lab frame (Lorentz) basis. We should underline that we claim there is a difference between $\vec{x}(t)$ and $\vec{x}_{cov}(t)$. In the main part of this paper we will investigate in detail the reason why this is the case, but we can give, already here,  a high-level, fundamental reason.   The trajectory $\vec{x}_{cov}(t)$ is viewed from the lab frame as the result of successive Lorentz transformations that depend on the proper time. In this  case relativistic kinematics effects arise. In view of the Lorentz transformation composition law, one will experience e.g. the Wigner rotation phenomenon. A Wigner rotation does not occur due to the action of forces, but has rather a pure kinematical origin.  In addition to the Wigner rotation one will also obtain other kinematics effects.  For instance, from the Lorentz transformations linking the instantaneous rest frames of the particle follows that the Einstein's rule of addition of velocities applies. In contrast to this, $\vec{x}(t)$ follows from the solution of the corrected Newton's equations and does not include relativistic kinematics effects: no Wigner rotation will be found, and the Galileo rule for adding velocities applies. Therefore, $\vec{x}(t)$ and $\vec{x}_{cov}(t)$ \textit{must} differ.

It is interesting to discuss the physical implications of this difference. We state that it depends on the choice of a convention, namely the synchronization convention of clocks in the lab frame. As such, it has no direct physical meaning. In fact, it is exactly the use of different synchronization conventions that yields different expressions for the particle trajectory in a single (e.g. the lab) reference system: different types of clocks synchronization certainly provide different time coordinates, but these coordinates describe the same reality. Note that without choosing a synchronization convention we cannot specify any experimental method by which simultaneity between two events in different places can be decided. In other words, the determination of simultaneous events imply the choice of a convention.
Whenever we have a theory containing an arbitrary convention, we should examine what parts of the theory depend on the choice of that convention and what parts do not. We may call the former convention-dependent, and the latter convention-invariant parts. Clearly, physically meaningful results must be convention-invariant.

The usual theoretical treatments of relativistic particle dynamics in the lab frame only involve Newton's second law corrected for the relativistic mass, and is based on the use of what we call "absolute time convention". This convention is a very natural choice, and this is the reason why this subject is never considered in conventional particle tracking calculations involved, for example, in accelerator and plasma physics. Covariant particle tracking is based, instead, on the use of Lorentz coordinates, and of a different synchronization convention, the Einstein's clocks synchronization convention.  Lorentz transformations show that the relation between space and time are not what is intuitively expected. Here, in contrast to the absolute time convention, we have a mixture of space and time.

There is a common mistake made in accelerator and plasma physics connected with the difference between $\vec{x}(t)$ and $\vec{x}_{cov}(t)$.
Let us look at this difference from the point of view of electrodynamics of relativistically moving charges. It is generally believed that the electrodynamics problem, similar to conventional particle tracking, can be treated within a description involving a single inertial frame. To evaluate fields arising from external sources we need to know their velocity and positions as a function of the lab frame time $t$. Suppose one wants to calculate properties of synchrotron (or cyclotron) radiation. Given our previous discussion the question arises, whether one should solve the usual Maxwell's equations in the lab frame with current and charge density created by particles moving along non-covariant trajectories like $\vec{x}(t)$. We claim that the answer to this question is negative, because non-covariant trajectories do not include relativistic kinematics effects.

In our previous work \cite{OURS1,OURS2,OURS3,OURS4,OURS5} we argued that the conventional algorithm for solving electromagnetic field equations, considered in all standard treatments as relativistically correct, is at odds with the principle of relativity. Many experts who learned relativistic dynamics and electrodynamics using standard textbooks (see e.g. \cite{LL,J}) will find this statement disturbing at first sight. However, our previous description implies quite naturally that usual Maxwell's equations in the lab frame are compatible only with the covariant trajectory $\vec{x}_{cov}(t)$, calculated by using Lorentz coordinates and, therefore, including relativistic kinematics effects.

\section{\label{sec:due} Relativistic kinematics and Galilean transformations}

It is generally accepted that in order to describe the dynamics of relativistic particles in the lab frame, which we assume inertial, one only needs to take into account the relativistic dependence of the particles momenta on the velocity. In other words, the treatment of relativistic particle dynamics involves only a corrected form of Newton's second law. To quote Feynman, Leiton and Sands \cite{F}: "Newton's second law, $d(m\vec{v})/dt = \vec{f}$, was stated with the tacit assumption that $m$ is a constant, but we now know that this is not true, and the mass of a body increases with velocity. (...) For those who want to learn just enough about it so they can solve problems, that is all there is to the theory of relativity - it just changes Newton's laws by introducing a correction factor to the mass."

According to this, the dynamics equation of a charged particle, e.g. an electron, in a fixed lab frame is

\begin{eqnarray}
	&& \frac{d\vec{p}}{dt} = e\left(\vec{E} + \frac{\vec{v}}{c}\times \vec{B}\right) ~,\cr &&
	\vec{p} = m\vec{v}\left(1 - \frac{v^2}{c^2}\right)^{-1/2}~ ,\label{N}
\end{eqnarray}
where rest mass, charge, and velocity are denoted by $m$, $e$, and $\vec{v}$  respectively. In a fixed lab frame, we consider an electric field $\vec{E}$ and a magnetic field $\vec{B}$. They interact with a charged particle in accordance with Eq.(\ref{N}). The Lorentz force law, plus measurements on the components of the acceleration of test particles, can be viewed as defining the components of electric and magnetic fields. Once the field components are found in this way, they can be used to predict the acceleration of other particles.

This study of relativistic particles motion looks precisely the same as in non relativistic Newtonian dynamics. Conventional particle tracking treats the space-time continuum in a non-relativistic format, as a (3+1) manifold. In other words, in the lab frame, Minkowski space-time "splits up" into three dimensional space and one dimensional time. This approach to relativistic particle dynamics relies on the use of three independent coordinates and velocities, "independent" meaning that there are no constraints among them. Once a prescribed force field is independently specified, the particle trajectory may be found by integration from initial conditions. The study a relativistic particle motion in a prescribed force field can thus be framed, mathematically, as a well-defined initial value problem.

Note that this solution of the dynamics problem in the lab frame makes no use of Lorentz transformations. This approach parallels non-relativistic ideas, and does not require the introduction of a four-dimensional Minkowski space. In fact, we do not have the typical mixture of positions and time that arises from Lorentz transformations. This way to deal with relativistic dynamics is commonly accepted in accelerator and plasma physics and echoing the previous quote from \cite{F}, it typically forces the physicist to believe that relativistic particle tracking is possible without detailed knowledge of the theory of relativity.

The velocity and acceleration  in Eq.(\ref{N}) are determined once the coordinates in the lab frame are chosen, and are measured at appropriate time intervals along the particle trajectory. However, in order to measure  the velocity of a particle within a single inertial lab frame, one first has to synchronize distant clocks.  While the concept of synchronization is very important in the understanding of special relativity, it seems that a clear exposition of it is lacking in literature: in particular, the type of clock synchronization which results in the time coordinate "$t$" in corrected Newton's equation Eq. (\ref{N}) is never discussed in accelerator and plasma physics.

We are now ready to discuss kinematics which is, in fact, a comparative study. It requires two relativistic observers and two coordinate systems\footnote{One might well wonder why it is necessary to discuss how different inertial frames are related to one another. The point is that all natural phenomena follow the principle of relativity, which  is a restrictive principle: it says that the laws of nature are the same (or take the same form) in all inertial frames. In agreement with this principle, usual Maxwell's equations can always be exploited in any inertial frame where electromagnetic sources are at rest. The fact that one can deduce electromagnetic field equations for arbitrary moving sources  by studying the form taken by Maxwell's equations under the transformation between rest frame of the source and the frame where the source is moving is a practical application of the principle of relativity.}. Since we require two coordinate systems, the question now arises how to assign a time coordinate to them.  The choice of convention on  clock synchronization is nothing more than a definite choice of coordinates system in an inertial frame of reference of the Minkowski space. In special relativity there are two practical choices of clock synchronization convention to consider:

(a) Einstein's convention,  leading to the Lorentz  transformations between frames

(b) Absolute time convention, leading to the Galilean transformations between frames

The dynamical evolution in the lab frame described by Eq. (\ref{N}) is based on the use of the lab frame time $t$ as independent variable, which is treated differently with respect to space. In other words, we want to solve the dynamics problem based on the use of the absolute time convention and we have no mixture of positions and time.  In this case the trajectory of a charged particle $\vec{x}(t)$, calculated by using the corrected Newton's second law, can be seen from the lab frame as the result of successive Galileo boosts that track the motion of the accelerated particle. As discussed in the introduction, the usual Galileo rule of addition of velocities is used to determine the  Galileo boosts tracking a particular particle along its motion.

Many physicists tend to think of Galilean transformations as pre-relativistic transformations between spatial coordinates and time that are not compatible with the special theory of relativity. However, this is not true. In the special theory of relativity, space-time is a flat, four-dimensional manifold with pseudo-Euclidean geometry. From this viewpoint, the principle of relativity is a simple consequence of the space-time geometry. The space-time continuum can be described in arbitrary coordinates, and the choice of this set of coordinates obviously cannot change the geometry of space-time.  Galilean transformations are simply transformations relating a given coordinate set to another coordinate set. As such they are actually compatible with the principle of relativity although, of course, they alter the form of Maxwell's equations.

\subsection{General form of pseudo-Euclidean metric}

Let us discuss in more detail how Galilean transformations can be understood in terms of the theory of special relativity. First, let us summarize a few well-known concepts. Any event in special relativity is mathematically represented by a point in space-time, called world-point. The evolution of a particle is, instead, represented by a curve in space-time, called world-line. If $ds$ is the infinitesimal displacement along a particle world-line, then

\begin{eqnarray}
	&& ds^2 =  c^2 dT^2 - dX^2 - dY^2 - dZ^2~ ,\label{M1}
\end{eqnarray}
where we have selected a special type of coordinate system (a Lorentz coordinate system),  defined by the requirement that Eq. (\ref{M1}) holds.

To simplify our writing we will use, instead of variables $T, X, Y, Z$,  variables $X^{0} = cT, ~ X^{1} = X,~ X^{2} = Y,~ X^{3} = Z$. Then, by adopting the tensor notation, Eq. (\ref{M1}) becomes $ds^2 = \eta_{ij}dX^{i}dX^{j}$, where Einstein summation is understood. Here $\eta_{ij}$ are the Cartesian components of the metric tensor and by definition, in any Lorentz system, they are given by $\eta_{ij} = \mathrm{diag}[1,-1,-1,-1]$, which is the metric canonical, diagonal form. As a consequence of the space-time geometry, Lorentz coordinates systems are connected by Lorentz transformations, which form the Lorentz group. Since the metric is invariant under Lorentz transformations the Lorentz group is also called the stability group of the metric.

The space-time continuum, determined by the interval  Eq. (\ref{M1}) can be described, however, in arbitrary coordinates and not only in Lorentz coordinates. In the transition to arbitrary coordinates the geometry of four-dimensional space-time obviously does not change, and in the special theory of relativity we are not limited in any way in the choice of a coordinates system. The space coordinates $x^1, x^2, x^3$ can be any quantities defining the position of particles in space, and the time coordinate $x^0$ can be defined by an arbitrary running clock.

The components of the metric tensor in the coordinate system $x^i$ can be determined by performing the transformation from the Lorentz coordinates  $X^{i}$ to the arbitrary variables $x^{j}$, which are fixed as $X^{i} = f^{i}(x^{j})$. One then obtains

\begin{eqnarray}
 && ds^2 = \eta_{ij}dX^{i}dX^{j} = \eta_{ij}\frac{\partial X^{i}}{\partial x^{k}}\frac{\partial X^{j}}{\partial x^{m}} = g_{km}dx^{k}dx^{m} ~ ,\label{M3}
\end{eqnarray}

This expression represents the general form of the pseudo-Euclidean metric.

\subsection{Space-time geometry and Einstein's postulates}

The derivation of relativistic kinematics is fairly elementary from a mathematical point of view, but it is conceptually subtle. Traditionally, the special theory of relativity is built on the principle of relativity and on a second additional postulate concerning the velocity of light:

1. The laws of nature are the same  (or take the same form) in all inertial frames

2. Constancy of the speed of light. Light propagates with constant velocity $c$ independently of the direction of propagation, and of the velocity of its source.


The basic point here is that the second ”postulate”, contrary to the view presented in textbooks, is not a physical assumption, but a convention that cannot be the subject of experimental tests. In fact, in order to measure the one-way speed of light one has first to synchronize the infinity of clocks assumed attached to every position in space, which allows us to perform time measurements. Obviously, an unavoidable deadlock appears  if one synchronizes the clocks by assuming a-priori that the one-way speed of light is $c$. In fact, in that case, the one-way speed of light measured with these clocks (that is the Einstein speed of light) cannot be anything else but $c$: this is because the clocks have been set assuming that particular one-way speed in advance. Therefore, it can be said that the value of the one-way speed of light is just a matter of convention without physical meaning. In contrast to this, the two-way speed of light, directly measurable along a round-trip, has physical meaning, because round-trip experiments rely upon the observation of simultaneity or non-simultaneity of events at a single point in space.


The theory of relativity can be deduced from postulates, as discussed above, but can also  be deduced  from the assumption of pseudo-Euclidean space-time geometry. In this case, dynamics and electrodynamics laws automatically include the principle of relativity and the postulate concerning the limiting character of the velocity of light. The difference between these two approaches is very interesting. However, as discussed above, assuming postulate 2 on the constancy of the speed of light in all inertial frames we also automatically assume Lorentz coordinates, meaning that Eq.(\ref{M1}) is valid, and that different inertial frames are related by Lorentz transformations. In other words, according to such limiting understanding of the theory of relativity, the fundamental invariant $d s^2$ equals the sum of the squares of the coordinates. This formulation of the theory of relativity is limited, because it assumes that only Lorentz transformations must be used to map the coordinates of events between inertial observers. In general, as we have seen, not only Lorentz coordinates and Lorentz transformations are permissible among inertial frames. Moreover, the invariant $ds^2$ for pseudo-Euclidean geometry, in an inertial frame with arbitrary coordinates obeys the more general relation Eq.(\ref{M3}).

\subsection{Pseudo-Euclidean metric and Galilean transformations}

Let us analyze some consequence of what we just discussed. We begin with the Minkowski metric as the true measure of space-time intervals for an inertial observer $S'$ with coordinates $(t',x')$. Here we neglect the two perpendicular space components that do not enter in our reasoning. We transform coordinates $(t,x)$ that would be coordinates of an inertial observer $S$ moving with velocity $-v$ with respect to the observer $S'$, using a Galilean transformation: we substitute $x' =x-vt$, while leaving time unchanged $t' = t$ into the Minkowski metric $ds^2 = c^2 dt'^2 - d x'^2$ to obtain

\begin{eqnarray}
&& ds^2 = c^2(1-v^2/c^2)dt^2 + 2vdxdt - dx^2  ~ .\label{G3}
\end{eqnarray}
Inspecting Eq. (\ref{G3}), or using transforming the Minkowski metric using the Galilean transformation above we can find the components of the metric tensor $g_{\mu\nu}$ in the coordinate system $(ct,x)$ of $S$. We obtain $g_{00} = 1-v^2/c^2$, $g_{01} = v/c$, $ g_{11} = - 1$. Note that the metric in Eq. (\ref{G3}) is not diagonal, since, $g_{01} \neq 0$, and this implies that time is not orthogonal to space.

This is a perfectly valid characterization of a moving inertial reference, see e.g. \cite{GENK}, Chapter 12,  for a discussion. For an observer co-moving with a particle at velocity $v$, $dx' = 0$. Hence we conclude that $ds^2 = c^2d\tau^2 = c^2(1-v^2/c^2)dt^2$, where $d\tau$ is the proper time, which is the time read off from a clock attached to the object. We note in passing that the quantity $g_{00} = (1-v^2/c^2)$, as we see from this formula, is positive, since $c > v$. If  this condition is not fulfilled,  the corresponding system of reference cannot be realized with real bodies.

In textbooks and monographs, the special theory of relativity is generally presented in relation to an interval $d s$ in the Minkowski form Eq.(\ref{M1}), while  Eq.(\ref{M3}) is ascribed to the theory of general relativity. To quote L. Landau and E. Lifshitz \cite{LL}:  "This formula is called the Galileo transformation. It is easily to verify that this transformation, as was to be expected, does not satisfy the requirements of the theory of relativity; it does not leave the interval between events invariant.".  This statement is obviously incorrect, because the space-time continuum can be described  equally well from the point of view of any coordinate system, which cannot possibly change $ds$. Authors of textbook \cite{LL} use the invariancy of $ds$, but  they understand it only in a limited sense when the metric is strictly diagonal \footnote{A comparison with three-dimensional Euclidean space  might help here. In the usual 3D Euclidean space, one can consider a Cartesian coordinate system ($x,y,z)$, a cylindrical coordinate system ($r,\phi, z$), a spherical coordinate system ($\rho, \theta, \phi$), or any other. Depending on the choice of the coordinate system one respectively has  $ds^2 = dx^2 + dy^2 + dz^2$, $ds^2 = dr^2 + r^2d\phi^2 + dz^2$, $ds^2 = d\rho^2 + \rho^2d\theta^2 + \rho^2\sin \theta^2d\phi^2$. The metric actually does not change, but the components of the metric do, depending only on the choice of coordinates. In general, in fact, we write $ds^2 = g_{ik}dx^idx^k$. Considering Cartesian coordinates, in particular, we will always have $g_{ij} = \mathrm{diag}(1,1,1)$.}.

The velocity of light in the coordinate system $(t', x')$ for $S'$, defined above as "at rest", is $c$. In the coordinate system $(t, x)$ \footnote{Transformation is interpreted in the passive sense.}, however, the speed of light cannot be equal $c$ anymore because $(t, x)$ is related to $(t',x')$ via a Galilean transformation. As a result, the speed of light in the direction  parallel to the $x$ axis is equal to $c + v$ in the positive direction, and $-c +v$  in the negative direction. This is readily verified if one recalls that the velocity of light in the reference system "at rest" is equal to $c$. If $ds$ is the infinitesimal displacement along the world line  of  a ray of light, then  $ds^2 = 0$ and we obtain $c^2 = (dx'/dt')^2$. In the moving reference system $x' = x - vt, ~ t' = t$ and this expression takes the form $c^2 = (dx/dt -v)^2$, which can be seen by a trivial change of variable, or setting $ds^2 =0$ in Eq. (\ref{G3}). This means that in the moving reference system of coordinates $(ct,x)$ the velocity of light parallel to the x-axis,  is $dx/dt = c + v$ in the positive direction, and $dx/dt = -c + v$  in the negative direction as stated above.

We conclude that the speed of light emitted by a  moving source measured in the lab frame $(t, x)$ depends on the relative velocity of source and observer, in our example $v$. In other words, the speed of light is compatible with the Galilean law of addition of velocities. The reason why it is different from the electrodynamics constant $c$
is due to the fact that the clocks are synchronized following the absolute time convention, which is fixed because $(t, x)$ is related to $(t',x')$ via a Galilean transformation. While most unusual in the theory of relativity, the choice of synchronizing clocks according to the absolute time convention is actually the most convenient in relativistic engineering. Note that from what we just discussed follows the statement that the difference between the speed of light and the electrodynamics constant $c$ is convention-dependent and has no direct physical meaning. In fact, the determination of simultaneous events implies the choice of a synchronization convention, and different types of clocks synchronization simply provide different time coordinates that describe the same reality. Similarly, in order to measure the speed of light, one first has to synchronize the clocks that measure the time interval as light travels between two given points in space. Therefore it can be said that, consistently with the conventionality of simultaneity, also the value of the speed of light is a matter of convention and has no definite objective meaning.

A widespread argument used to support the incorrectness of Galilean transformations is that they do not preserve the form-invariance of Maxwell's equations under a change of inertial frame. This idea is a part of the material in well-known books and monographs. To quote e.g. D. Bohm \cite{Bohm} "... the Galilean law of addition of velocities implies that the speed of light should vary with the speed of the observing equipment. Since this predicted variation is contrary to the fact, the Galilean transformations evidently cannot be the correct one.".  Similar statements can also be found in recently published  textbooks. To quote e.g. C. Cristodoulides \cite{Cr} "The fact that Galilean transformation does not leave Maxwell's equations has already been mentioned [...] On the other hand, experiments show that the speed of light in vacuum is independent of the source or observer.". These statements are incorrect: the constancy of the speed of light is related to the choice of synchronization convention, and cannot be subject to experimental tests.

Coordinates serve the purpose of labeling events in an unambiguous way, and this can be done in infinitely many different ways. The choice made in different cases is only a matter of convenience. In particular, as we have seen, absolute simultaneity can be introduced in special relativity without affecting the predictions of the theory.
In order for time to be absolute or global, it is necessary that all observers can agree that two events are simultaneous, whatever their respective relative velocities.  The absolute character of the temporal coincidence of two events is a consequence of the invariance of simultaneity $\Delta t' = \Delta t$ that follows from  Galilean transformations. As matter of fact, a coordinate system linked to the choice of absolute time synchronization is used in practice in accelerator and plasma physics because particle tracking calculations  becomes much simpler if the particle beam evolution is treated in terms of absolute time. This time synchronization convention is self-evident and this is the reason why this subject is not discussed in textbooks on relativistic engineering.

We would like to make some further remarks about kinematic relativistic effects. As discussed above, the Galilean transformation connecting the reference frame $S'$,  moving with velocity $v$ relative to the lab frame $S$, is given by  $x' =  x - vt, ~ t' = t$. This transformation implies a particular choice of synchronization convention  in the lab frame, which we called the "absolute time convention", so that the motion of particles looks precisely the same as predicted by Newtonian kinematics:  relativistic effects like Wigner rotation, time dilation,  Lorentz-Fitzgerald contraction and relativistic corrections in the law of composition of velocities  do not exist in this description.
In agreement with the principle of relativity, usual Maxwell's equations can be used in a moving inertial frame where a charge is instantaneously at rest. However, the transformation connecting any comoving frame to the lab frame in the case of the absolute time convention is a Galilean transformation, and Maxwell's equations do not remain invariant with respect to Galilean transformation. When a Galilean transformation of Maxwell's equations is tried, the new terms that have to be put into the Maxwell's equations lead to relativistic phenomena that were left out from the description of dynamics in terms of Newtonian kinematics. It does not matter which convention and hence transformation is used to describe the same reality. What matters is that, once fixed, such convention should be applied and kept in a consistent way in both dynamics and electrodynamics.

\subsection{Metric diagonalization}

The Galilean transformation connecting Lorentz coordinates $(t', x', y', z')$ with diagonal metric, Eq.(\ref{M1}), to coordinates $(t,x, y, z)$ with non diagonal metric, Eq.(\ref{G3}), is equivalent to a rotation in plane $x',t'$ with non-orthogonal axes $t,x$. In the coordinates system\footnote{As before, we can neglect $y$ and $z$.} $(t,x)$  we therefore have, as already discussed, much more complicated field equations. To get around this difficulty we observe that the non-diagonal metric can always be simplified. In fact, the space-time line-element in Eq.(\ref{G3})can be separated in a temporal part $d\tau$ and a spatial part $dl$ as

\begin{eqnarray}
	&& ds^2  = c^2 d\tau^2 - dl^2 ~ .\label{G6}
\end{eqnarray}

with

\begin{eqnarray}
	&& dl^2 = dx^2/(1 - v^2/c^2)  ~ .\label{G11}
\end{eqnarray}

and

 \begin{eqnarray}
 	&&  c^2d\tau^2 = \left[ \sqrt{1-v^2/c^2}cdt +  (v/c)dx/\sqrt{1-v^2/c^2}\right]^2 ~ .\label{G12}
 \end{eqnarray}

In practice we are "diagonalizing" the metric by completing the square and collecting terms in $dx$. Obviously, transforming to new variables leads to the usual Minkowski form of the metric. From Eq. (\ref{G6}) we find $dl/d\tau = c$. As expected, in the new variables the velocity of light is constant in all directions, and equal to the electrodynamics constant $c$. The overall combination of Galilean transform and variable changes specified above actually yields to the Lorentz transformation $dl = \gamma(dx' + vdt')~ , d\tau = \gamma(dt' + vdx'/c^2)$.

One should distinguish a coordinate velocity $dx/dt$ of a particle (or light) and its "physical" one. The latter is defined, see e.g. \cite{GENK} Chapter 2 for a definition, as the ratio of the "physical" distance $dl$ and the time interval $d\tau$ in the frame with Lorentz coordinates $(\tau, l)$.  We can conclude that in the case of Galileo boost from the particle's rest frame, coordinate  velocity of the particle $v$ coincides with "physical" velocity (Lorentz and Galilean transformations have the same parameter $v$). For any other coordinate velocity differing from $v$ there is no coincidence. In particular, for a light ray $dx/dt = v \pm c$ and $dl/d\tau = c$.  The discussion about the relation between coordinate quantities and "physical" quantities will be postponed until Section \ref{sec:sei}.

\subsection{Operational interpretation}

Let us give an "operational interpretation" of the results presented above. The fundamental laws of electrodynamics are expressed by Maxwell's equations, according to which, as well-known, light propagates with the same velocity $c$ in all directions. This is because Maxwell's theory has no intrinsic anisotropy. It has been stated that in their original form Maxwell's equations are only valid in inertial frames. However, Maxwell's equations can be written down in coordinate representation only if the space-time coordinate system has already been specified. The most natural assumption is that these equations are valid in Lorentz coordinates, i.e coordinates in which the metric has Minkowski form, Eq. (\ref{M1}). The decisive argument is that the constancy of the speed of light, which is a consequence of Maxwell's equations, is the basic properties of Lorentz coordinates attached to any inertial frame.

However, we have just discussed the fundamental importance that the issue of synchronization of  clocks in inertial frames has in the study of the motion of light sources. The question now arises how to assign space-time coordinates to an inertial lab frame where a source of light is at rest. We need to give a "practical", "operational" answer to this question. The most natural method of synchronization consists in putting all the ideal clocks together at the same point in space, where they can be synchronized. Then, they can be transported slowly to their original places (slow clock transport) \cite{CLOC}.

The usual Maxwell's equations are valid in any inertial frame where sources are at rest and the procedure of slow clock transport is used to assign values to the time coordinate. The same considerations apply when sources are moving in non-relativistic manner. In particular, when oscillating, charged particles emit radiation, and in the non-relativistic case, when charges oscillate with velocities much smaller than $c$, dipole radiation is generated and described with the help of the Maxwell's equations in their usual form. Since the value of the speed of light can be deduced from the Maxwell's equations, its constancy and isotropy is granted.

The theory of relativity offers an alternative procedure of clocks synchronization based on the constancy of the speed of light in all inertial frames. This is usually considered a postulate but, as we have seen, it is just a convention. The synchronization procedure that follows is the usual Einstein synchronization procedure.  Slow transport synchronization is equivalent to Einstein synchronization in inertial system where the light source is at rest. In other words, suppose we have two sets of synchronized clocks spaced along the $x$ axis. Suppose that one set of clocks is synchronized by using the slow clock transport procedure and the other by light signals. If we would ride together with any clock in either set, we could see that it has the same time as the adjacent clocks, with which its reading is compared. This is because in our case of interest, when light source is at rest, field equations are the usual Maxwell's equations and Einstein synchronization is defined in terms of light signals emitted by a source at rest assuming that light propagates with the same velocity $c$ in all directions.  Using any of these synchronization procedures in the rest frame we actually select a Lorentz coordinate system. In this coordinate system the metric has Minkowski form, Eq.(\ref{M1}).

We  now consider the case when the light source in the lab frame is accelerated from rest up to velocity $v$ along the $x$-axis. A fundamental question to ask is whether our lab clock synchronization method depends on the state of motion of the light source or not. The answer simply fixes a convention. The simplest method of synchronization consists in keeping, without changes, the same set of uniformly synchronized clocks used in the case when the light source was at rest, i.e. we still enforce the clock transport synchronization. This choice is usually the most convenient one from the  viewpoint of connection to laboratory reality.

It is clear that this synchronization convention preserves simultaneity and is actually based on the absolute time (or absolute simultaneity) convention. After the boost along the $x$ axis, the Cartesian coordinates of the emitter transform as $x' = x-vt, ~ y' = y, ~ z' = z$. This transformation completes with the invariance of simultaneity, $\Delta t' = \Delta t$. The absolute character of the temporal coincidence of two events is a consequence of the absolute concept of time, enforced by $t' = t$. As a result of the boost, the transformation of time and spatial coordinates of any event has the form of a Galilean transformation.  Since the emitter is at rest in the inertial frame $(t',x',y',z')$, it follows that, due to the principle of relativity, $(t',x',y',z')$ must be a Lorentz frame. Then, applying a Galilean transformation we obtain a non-orthogonal metric, Eq.(\ref{G3}), in $(t,x,y,z)$. We conclude accordingly that the metric, as well as the coordinate velocity of light from the moving emitter  in the lab frame is dependent on the relative velocity between emitter and observer and that the speed of light is compatible with the Galilean law of addition of velocities. Since clocks are synchronized by the absolute time convention, as we discussed in the previous sections, the velocity of light is different from the electrodynamics constant $c$. As before, the coordinate velocity of a ray of light parallel to the x-axis is given by $dx/dt = c + v$ in the positive direction, and $dx/dt = -c + v$  in the negative direction.

The fundamental point here is that, in agreement with the principle of relativity, the usual Maxwell's equations can always be exploited in a moving inertial frame where sources are at rest. However, the transformation connecting two inertial frames with absolute time synchronization is a Galilean transformation, and  Maxwell's equations do not remain form-invariant with respect to Galilean transformations. As a result, without changing the synchronization procedure in the rest frame, after the boost to the laboratory frame we have much more complicated electrodynamics of moving sources than usual. The main difference consists in the crossed terms $\partial^2/(\partial t\partial x)$, which arise from the d'Alambertian operator entering into the wave equation, due to the non-diagonal component of the metric tensor $g_{01} = v/c$. To get around this difficulty, we applied the diagonalization process described above. The trick needed here is to make a change of the time and spatial  variables. In the new variables Eq.(\ref{G11}) and Eq.(\ref{G12}) we obtain the metric in the usual Minkowski form, Eq.(\ref{G6}). Obviously, transforming to new variables leads to usual Maxwell's equations and we have standard electrodynamics of moving sources.

The question now arises how to operationally interpret these variable changes i.e. how one should change the rule-clock structure of the the lab reference frame. In order to assign a Lorentz coordinate system in the lab frame after the Galilean boost, one needs to perform additionally a distant clock  resynchronization $t \rightarrow t + xv/c^2$. After this, one needs to change the rhythm of all clocks $t \rightarrow \gamma t$, thus accounting for time dilation. The transformation of the rule-clock structure completes with the change scale of reference rules $ x \rightarrow \gamma x$, accounting for length contraction. This new space-time coordinates in the lab frame are interpreted, mathematically, by saying that the  metric is now diagonal and the speed of light from the moving source  is  isotropic and equal to $c$.

From an operational point of view, after the rule-clock structure transformation, the new coordinates in the lab frame are impeccable.  However, following  Einstein, the Lorentz time permits to perform the synchronization of clocks in the  lab reference system at different points in space with the aid of a light signal. The most important detail related with this procedure is the following: if the source of light is  in motion, the procedure for distant clocks synchronization in the lab frame must be performed by using the moving light source. The constant value of $c$ for the speed of light emitted by the moving source destroys the absolute (common) simultaneity and redefines the synchronization introduced by the slow clock transport procedure. The coordinates found by assuming a constant speed of light $c$ from a moving source are Lorentz coordinates for that particular source.

Consider now two light sources labelled with $1$ and $2$. Suppose that in the lab frame the velocities of the two sources are $v_1$, $v_2$ and $v_1 \neq v_2$. The question now arises, how to assign time coordinates to the lab reference frame. As discussed, we have a choice between absolute time coordinates and Lorentz time coordinates. The most natural pick, from the viewpoint of connection to the laboratory reality, is the absolute time  synchronization, related to the slow clock transport procedure. In this case simultaneity is absolute, and we only need one set of synchronized clocks in the lab frame, to be used for both sources. However, summarizing the previous observations, Maxwell's equations are not form-invariant under the Galilean transformations, that is, their form is different in the lab frame. Therefore, the use of the absolute time convention automatically implies the use of much more complicated field equations, and these equations are different for each source. One may therefore pick Lorentz time coordinates. The only possibility to introduce Lorentz coordinates in this situation consists in introducing individual coordinate systems  (i.e. individual set of clocks) for each source. It is clear that if operational methods are at hand to fix the lab coordinates for the first source, the same methods can be used to assign values to the coordinates for the second source and these will result in two different Lorentz coordinate systems and two different clock synchronizations in the lab frame. It should be clear that Lorentz coordinate systems are only a mental construct, but manipulations with non existing clocks are an indispensable prerequisite for the application of the usual Maxwell's equations for  moving light sources.

We now consider a relativistic particle accelerating in the lab inertial frame, and we analyze its evolution within the framework of special relativity, where the problem of assigning Lorentz coordinates to the lab frame in the case of acceleration motion is complicated. In fact, suppose that the particle is at rest in the lab frame for an instant. At this instant one picks Lorentz coordinates in the lab by using the clock transport synchronization. Then, in instant latter, the particle velocity changes to a infinitesimal value $v$ along the $x$-axis.  If the clock synchronization is fixed, one will have an infinitesimal Galilean transformation $x' = x- vt, ~ t' = t$ that describes the particle evolution. If the principle of relativity holds, however, $(t',x')$ must be a Lorentz frame. This is the central point of our argument. Therefore, in order to keep a Lorentz coordinates system in the lab frame, one needs to perform a clock resynchronization by introducing an infinitesimal time shift   $t' = t - xv/c^2$. This form of the Lorentz transformation is justified by the fact that we are dealing with an infinitesimal change in the particle velocity. Therefore, $v/c$ is so small that $v^2/c^2$ can be neglected, and one arrives at $x' = x - vt, ~ t' = t - xv/c^2$.   This infinitesimal Lorentz transformation just described differs from Galilean transformation only by the inclusion of the relativity of simultaneity, which is the only relativistic effect that appears in the first order in $v/c$.  All other higher order effects, that are Wigner rotation, Lorentz-Fitzgerald contraction, time dilation, and relativistic correction in the law of composition of velocities,  are derived mathematically, by iterating this infinitesimal transformation \cite{FA}.

\section{Connection between dynamics and kinematics}

Kinematics and dynamics are not independent  parts of physics. Some  four-vectors from kinematical treatments arise naturally in dynamics, such as the four-velocity and the four-acceleration. Kinematics studies the trajectory as a geometrical result independently of its cause. This means that it is not possible to predict the trajectory of a particle evolving under a given dynamical field using just a kinematical treatment.

Dynamics equations can be expressed as tensor equations in Minkowski space-time. When coordinates are chosen, one may work with components, instead of  geometric objects. Let us summarize some important points discussed before. Relying on the geometric structure of Minkowski space-time, one can define the class of inertial frames and can adopt a Lorentz frame with orthonormal basis vectors for any given inertial frame. Within the chosen Lorentz frame, Einstein's synchronization of distant clocks and Cartesian space coordinates are then automatically enforced, the metric tensor components are the usual $g_{\mu\nu} = \mathrm{diag}(1, -1, -1, -1)$, and any two Lorentz frames are related by a Lorentz transformation that preserves the metric tensor components, so that in any Lorentz coordinate system the law of motion becomes

\begin{eqnarray}
	&& m\frac{d^2 x_{\mu}}{d\tau^2} = e F^{\mu\nu}\frac{dx_{\nu}}{d\tau}~ ,\label{DE}
\end{eqnarray}
Here the electromagnetic field is described by the second-rank, antisymmetric tensor with components $F^{\mu\nu}$. The coordinate-independent proper time $\tau$ is a parameter describing the evolution of physical system under the relativistic laws of motion, Eq. (\ref{DE}).

The covariant equation of motion for a relativistic charged particle under the action of the four-force $K_{\mu} = e F^{\mu\nu}dx_{\nu}/d\tau$ in the Lorentz lab frame, Eq.(\ref{DE}), is a relativistic "generalization" of the Newton's second law. Its solution gives the world-line of the particles \footnote{The three-dimensional Newton second law $md\vec{v}/dt = \vec{f}$  can always be used in the instantaneous Lorentz comoving frame. Relativistic "generalization" means that these three independent equations are be embedded into the four-dimensional Minkowski space \cite{CT}. In Lorentz coordinates there is a kinematics constraint $u^{\mu}u_{\mu} = c^2$ for the four-velocity $u_{\mu} = dx_{\mu}/d\tau$. Because of this constraint, the four-dimensional dynamics law, Eq.(\ref{DE}),  actually includes only three independent equations of motion}.  Then one obtains a coordinate representation of the world-line in terms of lab ccordinates components of a four-vector describing, at any $\tau$, an event:  [$ct(\tau), x_1(\tau), x_2(\tau), x_3(\tau)$]. This position four-vector with components $x_{\mu}$ can be used to describe a particle's motion through space-time. To do so, we consider the spatial position as a function of the lab frame time $t$. This means that we form a world-line parametrized as  $x_{\mu} = [ct, \vec{x}_{cov}(t)]$.

When we  wish to describe the motion in the lab frame there is a method, alternative to that just described above, which parallels non-relativistic ideas. The dynamics of charge particles can be described by the conventional Lorentz force since the evolution parameter is, like in the non-relativistic case, the lab time $t$. We consider for this the spatial part of the dynamics equation, Eq.(\ref{DE}): $\vec{K} = (dt/d\tau) d(m\gamma\vec{v})/dt = \gamma d(m\gamma\vec{v})/dt$. The prefactor $\gamma$ arises from the change of the evolution variable from the proper time $\tau$, which is natural since $\vec{K}$ is the space part of a four-vector, to the lab frame time $t$, which is  needed to introduce the usual force three-vector $\vec{f}$: $\vec{K} = \gamma\vec{f}$. Written explicitly,  the relativistic form of the three-force is

\begin{eqnarray}
&& \frac{d(m\gamma\vec{v})}{dt} = e\left(\vec{E} + \frac{\vec{v}}{c}\times \vec{B}\right)~ .\label{DE1}
\end{eqnarray}
In other words, the law of relativistic dynamics within the Lorentz lab frame is very simple: aside for a straightforward correction in the relativistic mass, it looks precisely the same as in Newtonian dynamics. However, the simplicity of Eq.(\ref{DE1}) is to some extent an illusion.

In order to explain this statement, let us consider as in the last Section a relativistic particle accelerating in a lab inertial frame, and analyze its evolution within the Lorentz coordinate system. As before, suppose that the particle is at rest in the lab frame for an instant and that, at this instant one picks Lorentz coordinates. Then, an instant latter, the particle velocity changes to a infinitesimal value $dv$ along the $x$-axis. At this first step formula Eq.(\ref{DE1}) allows one to express differential $dv$ through the differential $dt$ in the rest Lorentz coordinate system. Then, in order to keep a Lorentz coordinate system in the lab frame, one needs to perform a clock resynchronization  by introducing the infinitesimal time shift   $t' = t - x(dv)/c^2$ - the only relativistic effect that appears in the first order.

It follows that in the lab Lorentz frame, during the next infinitesimal time interval, the  equation of motion  Eq.(\ref{DE1}) is valid after shifting the time axis. Then, instant after instant, the trajectory  $\vec{x}_{cov}(t)$ is viewed from the Lorentz lab frame as a result of successive infinitesimal Lorentz transformations. As we see, in Lorentz coordinates the lab time $t$ in the equation of motion Eq.(\ref{DE1}) cannot be independent from space variables.  This is because resynchronization of distant clocks in the process of particle acceleration leads to a mixture of positions and time.

In contrast to this, conventional particle tracking in accelerator and plasma physics treats the space-time continuum in a non-relativistic way as a (3 + 1) manifold \footnote{The fundamental difference is that conventional particle tracking is based on the same Eq.(\ref{DE1}) as covariant particle tracking, but the type of clock synchronization which provides time coordinate $t$ in the corrected Newton's equation is based on the use (much unusual in the theory of relativity) of the absolute time convention.}. Therefore, in this approach, one only introduces a modification of Newton's law by correcting for the relativistic mass, but time and space are treated differently. This approach does not require the introduction of Minkowski space-time and is widely used in the study of relativistic particle motion in prescribed force fields, since it is a well-defined initial value (Cauchy) problem \cite{RM}. Such approach to relativistic particle dynamics is based on the hidden assumption that the type of clock synchronization which provides the time coordinate $t$ in the lab frame relies on the use of the absolute time convention. The trajectory of the particle $\vec{x}(t)$, which follows from the solution of the corrected Newton's second law under the absolute time convention, does not include, however, relativistic effects \footnote{ Within the lab frame, if the particle motion follows the corrected Newton's second law  there cannot be Wigner rotation, Lorentz-Fitzgerald contraction, time dilation, and relativistic corrections in the law of composition of velocities. The introduction of the correction for the relativistic mass in the non-covariant equation of motion,  Eq.(\ref{DE1}), will be discussed in some detail later in (see section \ref{sec:5}).}.

We showed that $\vec{x}_{cov}(t)$ and $\vec{x}(t)$ differ from each other. Both are correct, and only refer to different choice of synchronization conventions. However, we criticize standard treatments of the coupling between electromagnetic sources and Maxwell's equations, since the trajectories in the source part of usual Maxwell's equations  are always identified with the trajectories obtained in the non covariant (3+1) manner. In other words, $\vec{x}(t)$ is always used, instead  of  $\vec{x}_{cov}(t)$ as it must be. We claim that  a solution of Maxwell's equations in their usual form based on the  results of conventional  particle tracking  $\vec{x}(t)$ cannot be used for the explanation of  experimental facts and that Maxwell's equations  are compatible, instead, with results of covariant particle tracking.

\section{Synchrotron radiation as an example}

In this section we discuss the emission of synchrotron radiation as an example illustrating the electrodynamics of moving bodies. A synchrotron radiation setup is a purely electrodynamic system where the source is a point charge moving  in the lab frame in a given constant magnetic field. Many of the features of this special problem are quite common in the theory of relativistic mechanics and electrodynamics. We consider a synchrotron radiation setup, because it is a physically interesting example for demonstrating that two different (Galilean and Lorentzian) synchronization procedures are possible in practice but with very different formulations in fundamental terms.

Let us consider the simple case when an ultrarelativistic electron moving along the $z$-axis in the lab frame is kicked by a weak dipole field directed along the $y$-axis before entering a uniform magnetic field directed along the $x$-axis, i.e. a bending magnet, and study the process of emission of synchrotron radiation in a bending magnet with and without the kick along the $y$-axis.

\subsection{Kicker setup description}

An electron kicker setup is a practical case of study for illustrating the difference between covariant and non-covariant trajectories.

Let us start with non-covariant particle tracking calculations. The trajectory of the electron, which follows from the solution of the corrected Newton's second law under the absolute time convention, does not include  relativistic effects. Therefore, as usual for Newtonian kinematics, Galilean vectorial law of addition of velocities is actually used. Non-covariant particle dynamics shows that the electron direction changes after the kick, while the speed remains unvaried. According to non-covariant particle tracking, the magnetic field $B\vec{e}_y$ is only capable of altering the direction of motion, but not the speed of an electron. This is clearly true when considering the equations of motion for a single electron  $dv_x/dt = \omega_c v_z$, $dv_z/dt = -\omega_c v_x$, where the characteristic "cyclotron" frequency $\omega_c$ is defined by $\omega_c = eB/(m\gamma )$. This is a well defined initial value problem with initial condition $v_z = v, v_x = 0, v_y = 0$.

In contrast, covariant particle tracking, which is based on the use of Lorentz coordinates, yields different results for the trajectory of the electron. Let us consider a setup with a relativistic electron and kicker, and analyze how a sequence of Lorentz boosts unfolds. Suppose an electron is moving at ultrarelativistic velocity $v$ parallel to $z$-axis, upstream the kicker. We assume for simplicity that the kick angle in the $x$-direction is small compared to $1/\gamma$. We further consider the small expansion parameter $\gamma v_x/c \ll 1$, neglecting terms of order $(\gamma v_x/c)^3$, but not of order $(\gamma v_x/c)^2$.

Let us consider a composition of Lorentz transformations that track the motion of the relativistic electron accelerated by the kicker field. Let the $S$ be a lab frame of reference and $S'$ a comoving with velocity $\vec{v}$ relative to $S$. Upstream the kicker, the particle is at rest in the frame $S'$. In order to have this, we impose that $S'$ is connected to $S$ by the Lorentz boost $L(\vec{v})$, with $\vec{v}$ parallel to the $z$ axis, which transforms a given four vector event $X$ in a space-time into $X' = L(\vec{v})X$. We study what happens in $S'$ before the kick. Our particle is at rest and the kicker is running towards it with velocity $-\vec{v}$. The moving magnetic field of the kicker produces an electric field orthogonal to it. When the kicker interacts with the particle in $S'$ we thus deal with an electron moving in the combination of perpendicular electric and magnetic fields.  It is easy to see that the acceleration in the crossed fields yields an electron velocity  $v'_x =\gamma v_x$ parallel to the $x$-axis and $v'_z = - v(\gamma v_x/c)^2/2$ parallel to the $z$-axis. If we neglect terms in $(\gamma v_x /c)^3$, the relativistic correction in the composition of velocities does not appear in this approximation.

Let $S"$ be a frame fixed with respect to the particles downstream the kicker. As is known, the composition of non collinear Lorentz boosts does not result in a simple boost but, rather, in a Lorentz transformation involving a boost and a rotation. In our second order approximation we can neglect the rotation. Therefore we can use a sequence of two commuting non-collinear Lorentz boosts linking $X'$ in $S'$ to $X''$ in $S''$ as $X"=L(\vec{e}_x v'_x) L(\vec{e}_z v'_z)X' =  L(\vec{e}_z v'_z)L(\vec{e}_x v'_x)X'$ in order to discuss the beam motion in the frame $S'$ after the kick. Here  $\vec{e}_x$ and $\vec{e}_z$ are unit vectors directed, respectively, along the $x$ and $z$ axis. Note that as observed by an observer on $S'$, the axes of the frame $S''$ are parallel to those of $S'$, and axes of $S'$ are parallel to those of $S$. The relation $X" = L(\vec{e}_x v'_x)L(\vec{e}_z v'_z)L(\vec{e}_z v)X$ presents a step-by-step change from $S$ to $S'$ and then to $S"$. For the simple case of parallel velocities, the addition law is $L(\vec{e}_z v'_z)L(\vec{e}_z v) = L(\vec{e}_z v_z)$. Here $v_z = v(1 - \theta^2/2)$ and $\theta = v_x/c$ in our case of interest. The resulting boost composition can be represented as $X" = L(\vec{e}_x v'_x)L(\vec{e}_z v_z)X$. This product of two non-collinear boosts is not a boost, but it can be represented as composition of a boost and a three-dimensional rotation: $L(\vec{e}_x v'_x)L(\vec{e}_z v_z) = R(\theta)L(\vec{n}v_z)$, where $R(\theta)$ is the rotation matrix of the system $S"$ through a Wigner angle $\theta = v_x/c$ in the $x,z$ plane of the system $S$, and $\vec{n}$ is the unit vector  $\vec{n} = \vec{e}_x\theta + \vec{e}_z (1-\theta^2/2)$. Note that we discuss particle tracking in the limit of a small kick angle $ \gamma v_x/c \ll 1$. However, even in this simple case and for a single electron we are able to demonstrate the difference between non-covariant and covariant particle trajectories.  The electron speed decreases from $v$ to $v(1-\theta^2/2)$. This result is at odds with the prediction from non-covariant particle tracking, because we used Lorentz transforms to track the particle motion. As a result, we track the particle in covariant way.

Let us now return to our consideration of the motion of a relativistic electron accelerated by the kicker field  and let us analyze the resynchronization process of the lab distant clocks during the acceleration of the electron. This will allow us to demonstrate a direct relation between the decrease of the electron speed after the kick in Lorentz coordinates and the time dilation phenomenon. As we already remarked, the Lorentz coordinate system is only a mental construction: manipulations with non existing clocks are only needed for the application of the usual Maxwell's equations for synchrotron radiation calculations.

Suppose that upstream the kicker we pick a Lorentz coordinates in the lab frame. Then, an instant  after entering the magnetic field, the electron velocity changes of the infinitesimal value $d\vec{v}$ along the $x$-axis. At this first step,  Eq.(\ref{DE1}) allows us to express the differential $d\vec{v}$ through the differential $dt$ in the Lorentz coordinate system assigned upstream the kicker. If clock synchronization is fixed, this is equivalent to the application of the absolute time convention. In order to keep Lorentz coordinates in the lab frame, as discussed before, we need to perform a clock resynchronization by introducing an infinitesimal time shift. The simplest case is when the kick angle $\theta$ is very small, and we evaluate transformations, working only up to the order $(\theta \gamma)^2$. The restriction to this order provides an essential simplicity of calculations in our case of interest for two reasons. First, relativistic correction to compositions of non-collinear velocity increments does not appear in this expansion order, but only in the order $(\gamma \theta)^3$. Second, the time dilation appears in the highest order we use.
Thus,  Eq.(\ref{DE1}) allows us to express the small velocity change $\Delta\vec{v}$ after the kick in the initial Lorentz coordinates system, and to perform clock resynchronization only downstream the kicker\footnote{In geometric language, Eq.(\ref{DE1}) is strictly verified on a hyperplane perpendicular to the electron world-line. The hyperplane tilts together with its normal (that is the four-velocity) $u_{\mu}$ as one moves along the particle world-line. As long as the velocity
increments are small enough and one can work only to the second order $(\gamma v_x/c)^2$, we can neglect the hyperplane tilt (i.e. the change of $u_{\mu}$), and the velocity increment can be calculated quite easily in the conventional way by integrating  Eq.(\ref{DE1}) and finding the velocity change without clock resynchronization (see Section \ref{sec:sei} for more details).}.

Therefore, after the kick we can consider the composition of two Lorentz boosts along the perpendicular $x$ and $z$ directions. The first boost  imparts the velocity  $v\theta\vec{e}_x$  to the electron along the $x$-axis and the second boost imparts the additional velocity $- (v\theta^2/2)\vec{e}_z$ along the $z$ axis, while the restriction to second order assures that the boosts commute. In order to keep a Lorentz coordinates system in the lab frame after the kick, that is equivalent to describe the kicker influence on the electron trajectory as Lorentz transformation, we need to perform a clock resynchronization by introducing a time shift and change the scale of time, that is the rhythm of all clocks, from $t$ to $\gamma_x t$, with $\gamma_x \simeq 1 + \theta^2/2$.  It is immediately understood that the speed of electron downstream the kicker is no longer independent of the electron motion in the magnetic field. No relativistic correction to the velocity component along the $x$-axis  appears in the second order, but a correction of the longitudinal velocity component, changing $v_z$ to $v_z/\gamma_x$ with $v_z = v(1 - \theta^2/2)$ and $v_z/\gamma_x = v(1 - \theta^2)$. It follows that the total electron speed in the lab frame, after  clock resynchronization downstream the kicker, decreases from  $v$ to $v(1 - \theta^2/2)$. It also turns out that the resulting Lorentz transformation contains a space rotation, i.e. a Wigner rotation, which is closely related with the time shift $t \rightarrow t - xv_x/c^2$, and to the time dilation $t \rightarrow \gamma_x t$, which leads to a relativistic correction of the longitudinal component of velocity. The time dilation does not come into the calculation of the velocity increment, but appears in the correction of the initial (relativistic) velocity $\vec{v} = v\vec{e}_z$.

This serious discrepancy between the results of non-covariant and covariant particle tracking naturally brings up the question:  which of these results is the correct one? The point is that both results describe correctly the same physical reality. The expressions for electron velocities are different only because they are based on the use of different synchronization conventions.  As we already mentioned above, the velocity of the electron has no definite objective meaning. In contrast to this, the direction of electron propagation downstream the kicker obviously has a direct objective meaning and does not depend on the choice of clock synchronization. The advantage of the absolute time convention is self-evident, since results in the applicability of Newtonian kinematics in the lab frame. In this case, the study of relativistic particle motion in a prescribed field is a well-defined initial value problem. The advantage of the standard clock synchronization, instead, is in the possibility of using Maxwell's equations in their usual form, yielding the solution of the electrodynamics problem with minimal efforts, compared to the case of the absolute time convention, which implies the use of much more complicated electrodynamics equations.

\subsection{Synchrotron radiation setup description}

An accelerated electron traveling on a curved trajectory emits radiation. When moving at relativistic speed, this radiation is emitted as a narrow cone tangent to the path of the electron. Moreover, the radiation amplitude becomes very large in this direction. This phenomenon is known as Doppler boosting. Synchrotron radiation is generated when a relativistic electron is accelerated in a bending magnet. Without going into the details of computation, it is possible to present intuitive arguments explaining why the characteristics of the spectrum  of synchrotron radiation only depend, in the ultrarelativistic limit,  on the difference between electron and light speed.

An electromagnetic source propagates through the system as a function of time following a certain trajectory $\vec{x}(t)$. However, an electromagnetic signal emitted at time $t$ at a given position $\vec{x}(t)$ arrives at the observer position at a different time $t_r$, due to the finite speed of light. As a result, an observer sees the  motion of the electromagnetic source as a function of $t_r$. Let us  discuss the case when the source is heading towards the observer. Using the fact that $c - v \ll c$ one obtains the well-known relation $dt_r/dt = (c - v\cos \phi)/c  \simeq (1 - v/c +\phi^2/2) \simeq (1/2)(1/\gamma^2 + \phi^2)$, where $\phi$ is the observation angle.  Since most of the emission takes place in the direction of motion of the electron within a cone of  opening angle $\phi \simeq 1/\gamma$ , the  observer perceives only the radiation emitted along a fraction of the trajectory of length $d_a \simeq R/\gamma$, where $R$ is the bending radius. For the observer, the duration of the signal emitted from this fraction of the trajectory is $T \simeq d_a/v -d_a/c = d_a(c-v)/(cv)$, so that in the ultrarelativistic limit under consideration $\omega_{cr} \simeq c\gamma^3/R$ is the characteristic frequency  of the synchrotron spectrum.

According to the correct coupling of fields and particles, there is a remarkable  prediction of synchrotron radiation theory  concerning the setup described above. Namely, there is a red shift of  the critical frequency of the synchrotron radiation in the kicked direction. To show this, let us first consider the covariant treatment, which makes explicit use of Lorentz transformations.

When the kick is introduced, covariant particle tracking predicts a non-zero red shift of the critical frequency, which arises because in Lorentz coordinates the electron velocity decreases from $v$ to $v - v\theta^2/2$, while the velocity of light is unvaried and equal to the electrodynamics constant $c$.
The red shift in the critical frequency can be expressed by the formula $\Delta\omega_{cr}/\omega_{cr} = - (3/2)\gamma^2 v_x^2/c^2 = - (3/2)\gamma^2\theta^2$. We now see a second order (time dilation) correction $\theta^2$ that is, however, multiplied by a large factor $\gamma^2$.

We can reinterpret the measurements with the help of Galilean transformations. According to non-covariant particle tracking the electron velocity is unvaried (i.e. there is no time dilation) but the velocity of light has increased from $c$, without kick, to $c(1 + \theta^2/2)$ with kick. This speed of light is compatible with the Galilean view: source and light velocities add up vectorially. When the kick is introduced we can write the total velocity of light in the lab frame  as the geometric sum of $c\vec{e}_z$ and $v_x\vec{e}_x$ (see section \ref{sec:due}). The reason for the velocity of light being different from the electrodynamics constant $c$ is due to the fact that, according to the absolute time convention, the clocks after the kick are not resynchronized.

We must mention another important point. In our  relativistic but non-covariant study of electron motion in a given magnetic field, the electron has  the same velocity and consequently the same relativistic factor $\gamma$ upstream and downstream the kicker. The motion in the bending magnet we obtained is practically the same as in the case of non-relativistic dynamics, the only difference being the appearance of the relativistic factor $\gamma$ in the determination of cyclotron frequency $\omega_c = eB/(m\gamma)$.  The curvature radius $R$ of the trajectory is derived from the relation $v_{\perp}/R  = \omega_c$, where $v_{\perp}$ is the component of the velocity normal to the field of the bending magnet $\vec{B} = B\vec{e}_x$. As a result, after the kick, the correction to the radius $R$  is only of order  $\theta^2$.

One could naively expect that according to covariant particle tracking the total speed of electron in the lab frame downstream the kicker decreases from $v$ to $v(1 - \theta^2/2)$, and that this would also lead to a consequent decrease of the 3D momentum from $m\gamma v$  to $m\gamma v(1 - \gamma^2\theta^2/2)$ in our approximation. However, a momentum change means a correction to the radius $R$ of order $\gamma^2\theta^2$ so that there is a glaring conflict with non covariant radius calculation. Since the  curvature radius of the trajectory in the bending magnet  has obviously an objective meaning, i.e. it is convention-invariant, this situation seems paradoxical. The paradox is solved taking into account the fact that  in Lorentz coordinates the 3D vector of momentum $\vec{p}$  is transformed, under Lorentz boosts, as the space part of the four vector $p_{\mu}$. Let us consider a composition of Lorentz boosts that track the motion of the relativistic electron accelerated by the kicker field. Under this composition of boosts the longitudinal momentum component remains unchanged in our approximation.

Let us verify that this assertion is correct. We have $p_{\mu} = [E/c,\vec{p}]$. We consider the Lorentz frame $S'$ fixed with respect to the electron upstream the kicker, and in the special case when electron is at rest $p_{\mu}' = [mc , \vec{0}]$. We turn focus on what happens in $S'$. Acceleration in the crossed kicker fields gives rise to an electron velocity $v_x' =  \gamma v_x$ parallel to the $x$-axis and $v_z' = - v(\gamma v_x)^2/2$ parallel to the $z$-axis. Downstream the kicker the transformed four-momentum is $p_{\mu}' = [mc + m v_x'^2/(2c), mv_x', 0, mv_z']$, where we evaluate the transformation only up to the order $(\gamma v_x/c)^2$, as done above. We note that, due to the transverse boost, there is a contribution  to the time-like part of the four-momentum vector i.e. to the energy of the electron. In fact, the energy increases
from $mc^2$ to $mc^2 + m (\gamma v_x)^2/2$. We remind that  $S'$ is connected to the lab frame $S$ by a Lorentz boost. Now, with a boost to a frame moving at velocity $\vec{v} = - v\vec{e}_z$, the transformation of the longitudinal  momentum component, normal to  the magnetic field of the bend, is $p_z = \gamma(p_z' + vp_0'/c) = \gamma mv$. Therefore we can see that the  momentum component along the $z$-axis remains unchanged in our approximation as it must be. We also have, from  the transformation properties of four-vectors,  that the time component $p_0 = \gamma(p'_0 + vp'_z) = \gamma mc$ .\footnote{It is clear (think for example to the non relativistic limit) that the energy-momentum four-vector is defined in terms of the particle's four-velocity as $p_{\mu} = mdx_{\mu}/d\tau = mu_{\mu}$. Actually we demonstrated above that by working only to order $(\gamma v_x/c)^2$, the four vector $u_{\mu}$ is subject to a constraint $du_{\mu}/d\tau = 0$ as one moves along a small fragment of the particle's world line. The vector relation Eq.(\ref{DE1}) is strictly valid on a hyperplane perpendicular to the world line which is tilted together with its normal $u_{\mu}$. Corrections to the four-velocity do not appear in the second order, but only in the order $(\gamma v_x/c)^3$. Thus, when it comes to the recovery of the particle evolution in our approximation,   Eq.(\ref{DE1}) allow us to  express the velocity increment $\Delta \vec{v}$ after the kick in the initial  Lorentz coordinate system (assigned upstream the kicker) and to perform clock resynchronization only downstream the kicker.}

Now everything fits together, and our calculations  show that covariant and non-covariant treatments give the same result for the red shift prediction, which is obviously convention-invariant and  has direct objective meaning. Only relative velocities play a role in synchrotron radiation phenomena. With the help of Lorentz transformations we could interpret the measurements of red shift as an evidence of the time dilation phenomenon. However, with the help of Galilean transformations we could interpret the same measurements of red shift by saying that the  speed of light from the moving source, measured in the lab frame, is dependent of the relative velocity between source and observer.

\subsection{Experimental test}

One way to demonstrate incompatibility between the standard approach to relativistic electrodynamics, which deals with the usual Maxwell's equations, and particle trajectories calculated by using non-covariant particle tracking is to make a direct laboratory test of synchrotron radiation theory. In other words, we are stating here that, despite the many measurements done during decades, synchrotron radiation theory is not an experimentally well-confirmed theory.

\subsubsection{Spontaneous emission}

In order to confirm the predictions of our coupling theory of fields and particles, we proposed in \cite{OURS2} a simple experiment at third generation synchrotron radiation sources with ultra-low electron beam emittance. Synchrotron radiation from bending magnets is emitted within a wide range of frequencies. The possibility of using narrow bandwidth sources in an experimental study on the red shift in synchrotron radiation spectrum looks more attractive. This allows one to increase the sensitivity of the output intensity on the red shift, and to relax the requirement on beam kicker strength and photon beam line aperture. Undulators, as sources of quasi-monochromatic synchrotron radiation, produce light in a sufficiently narrow bandwidth for our purposes. They cause the electron beam to follow a periodic undulating trajectory with the consequence that interference effects occur. Undulators have typically many periods. The interference of radiation produced in different periods results in a bandwidth that scales as the inverse number of periods. Therefore, the use of insertion devices installed at third generation synchrotron radiation facilities would allow us to realize a straightforward increase in the sensitivity to the red shift at a relatively small kick angle, $\theta < 1/\gamma$. The emittance of the electron beam in third generation synchrotron radiation sources is small enough to neglect finite electron beam size and angular divergence at least in the VUV wavelength range, and such synchrotron radiation source can be examined under the approximation of a filament electron beam. This allows us to take advantage of analytical presentations for single electron synchrotron radiation fields.  The spontaneous radiation pulse goes through a monochromator filter and its energy is subsequently measured by the detector. The proposed experimental procedure is relatively simple, because is based on relative  measurements in the velocity direction with and without transverse kick. Such a measurement is critical, in the sense that the prediction of conventional theory is the absence of the red shift, and has never been performed to our knowledge.

\subsubsection{Coherent emission}

There is another interesting problem where our correction of synchrotron radiation theory is required, which involves the production of coherent undulator radiation. Let us consider a microbunched ultrarelativistic electron beam kicked by a weak dipole field before entering a downstream undulator radiator. We want to study the process of emission of coherent undulator radiation from such setup. According to non-covariant particle tracking,  after the beam is kicked there is a trajectory change, while the orientation of the microbunching phase front remains as before. In other words, the kick results in a difference between the direction of the electron motion and the normal to the phase front. In standard electrodynamics, coherent radiation is emitted in the direction normal to the microbunching wavefront. Therefore, according to the conventional  coupling of fields and particles that we deem incorrect (i.e. according to usual algorithm for solving Maxwell's equations in the lab frame with charge and current density created  by particles moving along the trajectories calculated by using non covariant particle tracking), when the angular kick exceeds the divergence of the output coherent radiation, emission in the direction of the electron beam motion is strongly suppressed \cite{OURS6}. We have shown that our coupling of fields and particles predicts an effect in complete contrast to the conventional treatment \cite{OURS1,OURS2,OURS3,OURS4,OURS5}. Namely, when the evolution of the electron beam modulation is treated according to covariant particle tracking, the orientation of the modulation wavefront in the  ultra-relativistic asymptotic is always perpendicular to the electron beam velocity. In other words, relativistic kinematics shows the surprising effect that after the kick the orientation of the modulation wavefront is readjusted along the new direction of the electron beam. As a result, using standard electrodynamics we predict strong emission of coherent undulator radiation  from the  modulated electron beam in the kicked direction. It should be clear that in our example even the direction of emission of coherent undulator radiation is beyond the predictive power of the conventional synchrotron radiation theory. Such effect may be of practical importance in the analysis and interpretation of experiments with ultrarelativistic modulated electron beam at XFELs.

\section{\label{sec:5} The relativistic mass}

In the non covariant (3+1) space and time approach, there is no time dilation nor length contraction, because for Galilean transformations time and spatial coordinates scales do not change. Moreover, it can easily be verified that Newton's second law keeps its form under Galilean transformations. Therefore, in the (3+1) non covariant approach, there is  no kinematics correction factor $\gamma$ to the mass in Newton's second law. However, in contrast to kinematics effects like  time dilation and length contraction, the correction factor $\gamma$ to the mass in the Newton's second law  has direct objective meaning. In fact, if we assign space-time coordinates to the lab frame using the absolute time convention, the equation of motion is still given by Newton's second law  corrected for the relativistic dependence of momentum on velocity even though, as just stated, it has no kinematical origin. Understanding this result of the theory of relativity is similar to understanding previously discussed results: at first we use Lorentz coordinates and later the (3+1) non covariant approach in terms of a microscopic interpretation that must be consistent with the principle of relativity.

One can give a microscopic interpretation of the inertial mass of a particle as originating from the total  energy stored in the particle's field. In order to do so, we need to solve the  quantum field dynamics problem for our particle based on the use of quantities described in the non covariant approach. In agreement with the principle of relativity, and similarly to what we have discussed for electrodynamics, quantum field equations in their usual form can be used in a moving inertial frame where the particle is  at rest. However, the transformation connecting  comoving frame to the lab frame in the case of the absolute time convention is a Galilean transformation, and  field equations do not remain invariant with respect to Galilean transformations. We speculate that, similarly as for the electrodynamic problem, the new terms that have to be put into the field equations lead, in the lab frame, to the  correction factor $\gamma$ to the mass in the Newton's second law.

It is well-known from  classical electrodynamics that the electromagnetic field of an electron carries a momentum proportional to its velocity for $v \ll c$, while for an arbitrary velocity $v$, the momentum is altered by the relativistic $\gamma$ factor in the case when the absolute time convention is used. Many attempts have been made to explain the electron mass as fully originating from electromagnetic fields. However, these attempts have failed. In fact,  it is impossible to have a stationary non-neutral charge distribution held together by purely electromagnetic forces. In other words, mass and momentum of an electron cannot be completely electromagnetic in origin and in order to grant stability there is a necessity for compensating electromagnetic forces with non electromagnetic fields. From this viewpoint, Newton's second law is an empirical phenomenological law where the relativistic correction factor $\gamma$ to the mass is introduced in an ad hoc manner.

From a microscopic viewpoint, today accepted explanation of how structureless particles like leptons and quarks acquire mass is based on the coupling to the Higgs field, the Higgs boson having been recently experimentally observed at the LHC. This mechanism can be invoked to explain Newton's second law from a microscopic viewpoint even for structureless particles like electrons. However, at larger scales, an interesting and intuitive concept of the origin of physical inertia is illustrated, without recurring to the Higgs field, by results of Quantum Chromodynamics (QCD)  for protons and neutrons, which are not elementary and are composed of quarks and gluon fields. If an initial, unperturbed nuclear configuration is disturbed, the gluon field generates forces that tend to restore this unperturbed configuration. It is the distortion of the nuclear field that gives rise to the force in opposition to the one producing it, in analogy to the electromagnetic case. But in contrast to the electromagnetic model of an electron, the QCD model of a nucleon is stable,  and other compensation fields are not needed. Now, the gluon field mass can be computed  from the total energy (or momentum) stored in the field, and it turns out that the QCD version in which quark masses are taken as zero provides a remarkably good approximation to reality. Since this version of QCD is a theory whose basic building blocks have zero mass, the most of the mass of ordinary matter (more than 90 percent) arises from pure field energy \cite{W}. In other words, the mass of a nucleon can be explained almost entirely from a microscopic viewpoint, which automatically provides a microscopic explanation of Newton's second law of motion. In order to predict, on dynamical grounds, the inertial mass of a relativistically moving nucleon one does not need to have access to the detailed dynamics of strong interactions. It is enough to assume Lorentz covariance (i.e. Lorentz form-invariance of field equations) of the complete QCD dynamics involved in nucleon mass calculations \footnote{Lorentz covariance of the strong interactions is an unexplained fact, but all explanation must stop somewhere}. Then, the two  synchronization conventions discussed here give the same result for the relativistic mass correction, and it does not matter which transformation (Galilean or Lorentz) is used.

\section{\label{sec:sei}Discussion}

The laws of physics are invariant with respect to Lorentz transformations. This is a restrictive principle. Understanding the postulates of the theory of relativity is similar to understanding energy conservation: at first we learn this as a principle and later on  we study  microscopic interpretations that must be consistent with this principle. In the "microscopic" approach to relativistic phenomena,  Lorentz covariance of all the fundamental laws of physics remains, similarly to energy conservation, an unexplained fact.

There is no single right way to think about relativistic phenomena. For example, in the non covariant (3+1) space and time approach, there is no time dilation, since for Galilean transformations the time scale does not change. It is therefore natural to ask, whether time dilation is real, or just  a mathematical trick. The answer is sophisticated. In a Lorentz coordinate system (i.e. using standard clock synchronization)  time dilation is a real phenomena, in sense that we can present an experimental setup  where the outcome depends on the change in the rhythm of a moving clock. Think, for example, of the muon decay. However, as already discussed, there is another satisfying  way to describe the same experimental setup based on the absolute time convention, and in this case the usual quantum field equations, describing the dynamics, cannot be employed. The new terms that have to be put into the field equations due to the use of Galilean transformations lead to the same prediction as concerns experimental results: the muon population in the lab frame, after the travel distance $v\tau_0/\sqrt{1-v^2/c^2}$, is reduced to 1/2 of the origin population, where $\tau_0$ is the half-life at rest.
The two approaches give the same  result for the travel distance, which is convention-invariant, and it does not matter which  transformation (Galilean or Lorentz) is used: they both describe the same reality.

It is important to stress at this point that the dynamical line of argument explains what the Minkowski geometry physically means. The pseudo-Euclidean geometric structure of space-time is only an interpretation of the behavior of the dynamical matter fields in the view of different observers, which is an observable, empirical fact. It should be clear that the relativistic properties of the dynamical matter fields are fundamental, while the geometric structure is not. For example, muons in motion behave relativistically because the field forces that are responsible for the muon disintegration satisfy equations that are Lorentz covariant. Dynamics, based on the quantum field equations, is actually hidden in the language of kinematics. The Lorentz covariance of the equations that govern the fundamental interactions of nature is an empirical  fact, while the postulation of the pseudo-Euclidean geometry of space-time is a mathematical interpretation of it that yields the laws of relativistic kinematics, but is in fact based on the way the field behave dynamically.

\subsection{Geometric restatement of Newton's second law}

Let us now return to the Lorentz transformations and try to get a better understanding of the  geometric restatement  of Newton's second law. To derive the covariant form of relativistic dynamics, we should embed the three-dimensional vector relation $md\vec{v}/dt = \vec{f}$ into the four-dimensional geometry of Minkowski space \cite{CT}.  The idea of embedding is based on the principle of relativity i.e. on the fact that the usual Newton's second law can always be used in any Lorentz frame where the particle, whose motion we want to describe, is at rest. In other words, if an instantaneously comoving Lorentz frame is given at some instant, one can precisely predict the evolution of the particle in this frame during an infinitesimal time interval. In geometric language, the vector relation Eq.(\ref{DE1}) is strict on a hyperplane perpendicular to the world line. However, the hyperplane tilts together with its normal $u_{\mu}$ as one moves along the world line. For the embedding we need an operator $\hat{P}_\perp$ that continually projects vectors of Minkowski space on hyperplanes perpendicular to the world line. The desired operator is $(\hat{P}_\perp)_{\mu\nu} = \eta_{\mu\nu} - u_{\mu}u_{\nu}/u^2$ \cite{CT}. In the instantaneously comoving frame one can unambiguously construct a four-force $K_{\mu} = [0,\vec{f}]$. Then, in an arbitrary Lorentz inertial frame, the components  $K_{\mu}$ can be found through the appropriate Lorentz transformation. In the rest frame obviously $u_{\mu}K^{\mu} = 0$. It follows that, since $u_{\mu}K^{\mu}$ is an invariant, the four-force $K_{\mu}$ is perpendicular to the four-velocity $u_{\mu}$ in any Lorentz frame. The desired embedding of Newton's second law in hyperplanes perpendicular to the world line is found by imposing $(\hat{P}_\perp)_{\mu\nu}(m du_{\nu}/d\tau - K_{\nu}) = 0$.  This is a tensor equation in Minkowski space-time that relates geometric objects and does not need coordinates to be expressed. The evolution of a particle can be described in terms of world line $\sigma(\tau)$, and the 4-velocity by $u = d\sigma/d\tau$, having a meaning independently of any coordinate system. Similarly, in geometric language, the electromagnetic field is described by the second-rank, antisymmetric tensor $F$, which also requires no coordinates for its definition. This tensor produces a 4-force on any charged particle given by $\hat{P}_\perp\cdot(mdu/d\tau - eF\cdot u) = 0$ \cite{CT}. This is the basic dynamics law for relativistic charged particles expressed in terms of geometric objects and automatically includes the principle of relativity. The presence of the projector operator $\hat{P}_\perp$ suggests that we have only three independent equations. In the case of Maxwell's equations we are able to rewrite the equations in the relativistic form without any change in the meaning at all, just with a change notation. It is important to notice that the situation with dynamics equation is more complicated.

In order fully to understand the meaning of the embedding of the dynamics law in the hyperplanes perpendicular to the world line, one must keep in mind that, above, we  characterized the Newton's equation in the Lorentz comoving frame as a phenomenological law. The microscopic interpretation of the inertial mass of a particle is not given. In other words, it is generally accepted that Newton's second law is an phenomenological law  and the rest mass is introduced in an ad hoc manner. The system of coordinates in which the equations of Newton's mechanics are valid can be defined as Lorentz rest frame.  The relativistic generalization of the Newton's second law to any Lorentz frame permits us to make correct predictions. The projector operator guarantees that this coordinate system restriction will be satisfied. In the non covariant (3+1) approach there is no kinematics correction factor $\gamma$ to the mass in Newton's second law, and the relativistic dependence of the particles momenta on the velocity in the case of non-standard clock synchronization can be interpreted only on dynamical grounds (or accepted as an empirical dependence). In the preceding section we pointed that, according to the present level of understanding, the mass of a nucleon comes from the total energy stored in the nucleon gluon field.  The strong interaction field equations are Lorentz covariant and can be expressed (similarly to electromagnetic field equations) as  equations relating geometric objects in  space-time. To simplify computations one can work with arbitrary coordinate formulation of field equations. The choice between these different possibilities is a matter of pragmatics. Today we know that there is a machinery behind the nucleon's inertial mass. It origin is explained in framework of the Lorentz covariant quantum field theory. In the microscopic approach to the inertial mass,  Einstein and absolute time   synchronization conventions give the same result for any convention-invariant phenomena, and it does not matter which transformation (Galilean or Lorentz) is used. The dynamical line of arguments explains a situation when the relativistic mass correction cannot be interpreted as a relativistic kinematic effect in the (3+1) (i.e. separate three-dimensional space and one-dimensional time) coordinatization.

\subsection{Phenomenology and relativistic extensions}

We are in the position to formulate the following general statement: any phenomenological law, which is valid in the Lorentz rest frame, can be embedded  in the four dimensional space-time only by using Lorentz coordinatization (i.e. Einstein synchronization convention). Suppose we do not know why the muon disintegrates, but we know the law of decay in the Lorentz rest frame. This law would then be a phenomenological law. The relativistic generalization of this law to any Lorentz frame allows us to make a prediction on the average distance travelled by a muon. In particular, when a Lorentz transformation of the decay law is tried, one obtains the prediction that after the travel distance $\gamma v\tau_0$, the population in the lab frame  would be reduced to 1/2 of the origin population. We may interpret this result by saying that, in the lab frame, the characteristic lifetime of a particle has increased from $\tau_0$ to $\gamma\tau_0$. In contrast, in the non covariant (3+1) space and time approach there is no time dilation effect, since for Galilean transformations the time scales do not change.  Therefore,   in the (3+1) non covariant approach, there is no kinematics correction factor $\gamma$ to the travel distance  of relativistically moving muons. The two approaches give, in fact, a different result for travel-distance, which must be, however, convention-invariant.  This glaring conflict between results of covariant and non covariant approaches is explained similarly to the case of relativistic mass correction treated above: it is a dynamical line of arguments that explains this paradoxical situation with the relativistic $\gamma$ factor. In fact, there is a machinery behind the muon disintegration. Its origin is explained in the framework of the Lorentz-covariant quantum field theory. In the microscopic approach to muon disintegration,  Einstein and absolute time   synchronization conventions give the same result for such convention-invariant observables like the average travel distance, and it does not matter which transformation (Galilean or Lorentz) is used.

\subsection{Galilean transformations and electromagnetic field theory}

Let's now go back to our calculations of the speed of light from a moving source when the clocks in the lab frame are synchronized by the absolute time convention. Back in Section 2, we found that, due to this particular choice of synchronization convention in the lab frame, the speed of light is compatible with the Galilean law of addition of velocities. We used four-geometric arguments (i.e. the  language of relativistic kinematics) to show that in the lab frame, where the source is moving with velocity $v$ along the $x$-axis, the velocity of light in the direction parallel to the $x$-axis, is equal to $c+v$ in the positive, and $-c+v$  in the negative orientation. For many practical purposes, it will still be convenient to use a four-geometric approach. However, we are now ready to study the same outcome in terms of properties of the dynamical fields. In fact, light propagation  can  be explained in the framework of the electromagnetic field theory.

In the comoving frame, fields are expressed as a function of the independent variables $x', y', z'$, and $t'$. Let us consider Maxwell's equations in free space.  The electric field $\vec{E}'$ of an electromagnetic wave satisfies the equation $\Box'^2\vec{E}' =  \nabla'^2\vec{E}' - \partial^2\vec{E}'/\partial(ct')^2  = 0$.
However, the variables $x',y',z',t'$ can be expressed in terms of the independent variables $x, y, z, t$ by means of a Galilean transformation, so that fields can be written in terms of  $x, y, z, t$. From the Galilean transformation $x' = x - vt, ~ y' = y, ~ z' = z, ~ t' = t $, after partial differentiation, one obtains

\begin{eqnarray}
	\frac{\partial }{\partial t} = \frac{\partial }{\partial t'} - v \frac{\partial }{\partial x'}~ , ~
	\frac{\partial }{\partial x} = \frac{\partial }{\partial x'} ~.
\end{eqnarray}

Hence the wave equation transforms into

\begin{eqnarray}
	&& \Box^2\vec{E} = \left(1-\frac{v^2}{c^2}\right)\frac{\partial^2\vec{E}}{\partial x^2}  - 2\left(\frac{v}{c}\right)\frac{\partial^2\vec{E}}{\partial t\partial x}
	+ \frac{\partial^2\vec{E}}{\partial y^2} + \frac{\partial^2\vec{E}}{\partial z^2}
	- \frac{1}{c^2}\frac{\partial^2\vec{E}}{\partial t^2} = 0 ~ \label{GT}
\end{eqnarray}

where coordinates and time are transformed according to a Galilean transformation. The solution of this equation is the sum of two arbitrary functions, one of argument $x - (c +v)t$ and the other of argument $x + (-c +v)t$:

\begin{eqnarray}
F[x - (c +v)t] + G[x + (-c +v)t] ~.
\end{eqnarray}

Here we obtained the solution for waves which move in the $x$ direction by supposing that the field did not depend on $y$ and $z$. The first term represents a wave traveling forward in the positive $x$ direction, and the second term a wave traveling backwards in the negative $x$ direction. This result agrees with what we would have found more rapidly using the metric Eq.(\ref{G3}).  However, in this way we have  provided a dynamical underpinning for our previous discussion of the behavior of the speed of light under a Galilean transformation.

In section 2 we already found that, starting from the diagonal form of the metric tensor in the rest frame and applying a Galilean transformation we obtain the non-diagonal metric Eq.(\ref{G3}). We observed that this non-diagonal metric can always be simplified. In particular, we could transform it to the usual Minkowski form by changing variables. Let us take the dynamical field viewpoint and use it to understand this change of variables.

After properly transforming the d'Alembertian through a Galileo boost, which changes the initial coordinates $(x',y',z',t')$ into $(x,y,z,t)$, we can see that the homogeneous wave equation for the field in the lab frame  has nearly but not quite  the usual, standard form that takes when there is no uniform translation in the transverse direction with velocity $v$. The main difference consists in the crossed term $\partial^2/\partial t\partial x$, which complicates the solution of the equation. To get around this difficulty, we observe that simplification is always possible. The trick needed here is to further make a change of the time variable according to the transformation $t' = t - x v_x/c^2$. In the new variables in i.e. after the Galilean coordinate transformation and the time shift we obtain the  d'Alambertian in the following form \footnote{It should be clear that, in principle, the transformed wave equation may be solved directly without  change of variables, for example by numerical methods, and one may directly derive physical (i.e. convention-invariant) effects associated with the "crossed" term.}

\begin{eqnarray}
\Box^2 =	\left(1-\frac{v_x^2}{c^2}\right)\frac{\partial^2}{\partial x^2} + \frac{\partial^2}{\partial y^2} + \frac{\partial^2}{\partial z^2}
- \left(1-\frac{v_x^2}{c^2}\right)\frac{1}{c^2}\frac{\partial^2}{\partial t^2} ~.
\end{eqnarray}
A further change of a factor $\gamma$  in the scale of time and of the coordinate along the direction of uniform motion leads to the usual Maxwell's equations. In particular, when coordinates and time are transformed according to a Galilean transformation followed by the  variable changes specified above, the  d'Alambertian  $\Box'^2 = \nabla'^2 - \partial^2/\partial(ct')^2$  transforms into  $\Box^2 = \nabla^2 - \partial^2/\partial(ct)^2$ . The overall combination of Galileo transformation and variable changes actually yields the Lorentz transformation in the "3+1" space and time

\begin{eqnarray}
x' = \gamma(x -  v_x t), ~ y' = y, ~ z' = z, ~ t' = \gamma(t - x v_x/c^2)  ~, \label{LTT31}
\end{eqnarray}
but in the present context Eqs. (\ref{LTT31}) are only to be understood as useful mathematical devices, which allow one to solve the electrodynamic problem in the "3+1" space and time with minimal effort.

Since the Galilean transformation, completed by the introduction of the new variables,  is mathematically equivalent to a Lorentz transformation, it obviously follows that transforming to new variables leads to the usual Maxwell's equations.

We state that the variable changes performed above have no intrinsic meaning - their meaning only being assigned by a convention. In particular, one can see the connection between the time shift $t' = t - xv_x/c^2$ and the issue of clock synchrony. Note that the final change in the scale of time and spatial coordinates is also unrecognizable from a physical viewpoint. It is clear that  the convention-independent results of calculations  are precisely the same in the new variables. As a consequence, we should not care to transform the results of the electrodynamics problem solution  into  the original variables.

\subsubsection{The non-relativistic limit}

It is generally believed that a Lorentz transformation reduces to a Galilean transformation in the non-relativistic limit. Let us present a typical textbook statement \cite{FR} concerning the non relativistic limit of Lorentz transformations:  "The reduction of $t'= \gamma(t - vx/c^2)$ to Galilean relation $t' = t$ requires $x \ll ct$ as well as $v/c \ll 1$". We state that this is incorrect and misleading. As discussed, kinematics is a comparative study which requires two coordinate systems, and one needs to assign time coordinates to the two systems. Different types of clock synchronization provide different time coordinates.

The convention on the clock synchronization amounts to nothing more than a definite choice of the coordinate system in an inertial frame of reference in Minkowski space. Pragmatic arguments for choosing one coordinate system over another may therefore lead to different choices in different situations. Usually, in relativistic engineering, we have a choice between absolute time coordinate and Lorentz time coordinate. The space-time continuum can be described equally well in both coordinate systems. This means  that  for arbitrary particle speed,  the Galilean coordinate transformations well characterize a change in the reference frame from the lab inertial observer to a comoving inertial observer in the context of the theory of relativity. Let us consider the non relativistic limit. The Lorentz transformation, for $v/c$ so small that $v^2/c^2$ is neglected can be written a $x' = x - vt$, $t' = t - xv/c^2$. This infinitesimal Lorentz transformation differs from the infinitesimal Galilean transformation $x' = x - vt$, $t' = t$. The difference is in the term $xv/c^2$ in the Lorentz transformation for time, which is a first order term. If the lab frames $S$ and comoving frame $S'$ have coordinates in a non standard (absolute time) configuration one needs to transform Maxwell's equations according to a Galilean transformation, and we obtain Eq.(\ref{GT}). We can see that the wave equation in the lab frame after the Galileo boost has non-diagonal form even in the non-relativistic limit $v/c \ll 1$, $\gamma \sim 1$. The difference consists in the crossed term $\partial^2/\partial t\partial x$ which arises when applying the Galileo boost.

Let us consider the non relativistic limit  in the context of the effect of light aberration, that is a change in the direction of light propagation ascribed to boosted light sources. The explanation of the effect of aberration of light presented in well-known textbooks is actually based on the use of a Lorentz boost (i.e. of relativistic kinematics) to describe how the direction of a light ray depends on  the velocity of the light source relative to the lab frame. Let us discuss a special case of the aberration of a horizontal light ray. Suppose that a light source, studied in the comoving frame $S'$, radiates a plane wave along the $z$-axis. Now imagine what happens in the lab frame,  where the source  is moving with constant speed $v$ along the $x$-axis. The transformation of observations from the lab frame  with Lorentz coordinates to the comoving Lorentz frame is described by a transverse  Lorentz boost. On the one hand, the wave equation remains invariant with respect to Lorentz transformations. On the other hand, if make a Lorentz boost, we automatically introduce a time transformation $t' = t - xv/c^2$ and the effect of this transformation is just a rotation of the wavefront in the lab frame. This is because the effect of this time transformation is just a dislocation in the timing of processes, which has the effect of rotating the plane of simultaneity on the angle $v/c$ in the first order approximation. In other words, when a uniform translational motion of the source is treated according to Lorentz transformations, the aberration of light effect is described in the language of relativistic kinematics. In fact, the relativity of simultaneity is a  relativistic effect that appears also in the first order in $v/c$.

It should be noted, however, that there is another satisfactory way of explaining  the effect of aberration of light. The explanation consists in using a Galileo boost to describe the  uniform translational motion of the light source in the lab frame. After the Galilean transformation of the wave equation we come to the conclusion that the crossed term described above yields an aberration angle $v/c$. In fact,  in order to eliminate the crossed term in the transformed wave equation, we make a change of the time variable. After Galilean coordinate transformation and time shift we obtain the wave equation in "diagonal" form, i.e. without crossed terms. The time shift results in a slope  of the plane of simultaneity. Then, the electromagnetic waves are radiated at the angle $v/c$, yielding the phenomenon of light aberration: the two approaches, treated according to Einstein's or absolute time synchronization conventions give the same result. The choice between these two different approaches is a matter of pragmatics. However, we would like to emphasize a difference in the conceptual background between these two approaches. The non-covariant (Galilean) approach gives additionally a physical insight into the particular laws of nature it deals with. For instance, the dynamical line of arguments explains the aberration of light based on the structure of the electromagnetic field equations, even if hidden in the language of relativistic kinematics (relativity of simultaneity).

\subsection{Relativistic kinematics effects and ultrarelativistic asymptotic}

We now consider again the kicker setup described in section 4, in order to present an analysis of how the various relativistic effects turn up in successive orders of the small kick angle. Before turning to the actual discussion, we mention briefly  the case when the particle is accelerating from  rest. The appearance of a relativistic effects does not depends on the use a large relative speed of the two reference frames. Suppose that the particle is at rest in an inertial frame for an instant. At this instant, one picks a Lorentz coordinates system. Then, an instant later, the particle velocity changes of an infinitesimal value $dv$ along the $x$-axis.  The Lorentz transformation describing this change, for $dv/c$ so small that $dv^2/c^2$ is neglected, is described by $x' = x - ct(dv/c), ~ ct' = ct - x(dv/c)$.   The relativity of simultaneity is, then, the only relativistic effect that appears in the first order in $dv/c$.   To obtain a transformation valid for a finite relative speed between two reference frames, we must consider $n$ successive infinitesimal transformations, and then take the limit $n \longrightarrow \infty$, $dv/c \longrightarrow 0$, $ndv/c  \longrightarrow v$.
Consider first the case in which $v/c$ is fairly small, so that we neglect $v^3/c^3$, but not $v^2/c^2$. This case yields effects of the second order, which need to be considered in addition to the relativity of simultaneity, which appears already in the first order. Also time dilation and length contraction appear in the second order $v^2/c^2$, while the relativistic correction in the composition of velocities only appears in the order $v^3/c^3$ and higher. If the increments of the velocity are not all in the same direction, the transformation matrices do not commute, and this originates a Wigner rotation, which also appears in the order $v^2/c^2$.

Let us now turn to the kicker setup, and see what happens in the situation when a modulated ultrarelativistic electron beam, moving along the $z$-axis, is kicked by a weak dipole field  directed along the $y$-axis. Consider  the case in which $\gamma v_x/c$ is fairly small, so that we neglect $\gamma^3v^3/c^3$, keeping however terms in $\gamma^2v^2/c^2$. Suppose that upstream the kicker we pick a Lorentz coordinates system in the lab frame. In order to keep the Lorentz coordinate system in the lab frame after the kick we need to describe the kicker influence on the electron trajectory as a result  of successive  Lorentz transformations. The restriction to the second order  $\gamma^2v^2/c^2$ yields simple calculations. Namely, after the kick, we can consider only the composition of two Lorentz boosts along perpendicular to the $x$ and $z$ directions. The first transverse boost imparts the velocity $v\theta\vec{e}_x$  and the second boost the additional velocity $- (v\theta^2/2)\vec{e}_z$ along the $z$-axis. The restriction to the second order allows for  boost commutation. It is important to study the influence of the transverse boost. The boost along the direction of the $x$-axis is given by the transformation  $t' = \gamma_x(t - x v_x/c^2)$, $y' =y, z' = z$,  $x' = \gamma_x(x - tv_x)$. In the second order approximation, the factor $\gamma_x = 1/\sqrt{1-v_x^2/c^2}$ approximates to $\gamma_x \simeq 1 + \theta^2/2$.

Let us see how the Wigner rotation appears. Suppose the beam velocity is perpendicular to the wavefront of the modulation upstream the kicker. Seen from the lab frame, the wavefront of the beam modulation rotates relative to  the Cartesian axes of the lab frame when a modulated electron beam is accelerated in the kicker field. The expression for an infinitesimally small rotation angle  is given by \cite{Rit}

\begin{eqnarray}
\delta \Phi = 	\left(1 - \frac{1}{\gamma} \right)\frac{\vec{v}\times d\vec{v}}{v^2}	
=	\left(1 - \frac{1}{\gamma} \right) \delta \theta ~.
\label{d}
\end{eqnarray}
where $d\vec{v}$ is the vector of an infinitesimally small  velocity change due to acceleration, $\Phi$  is the Wigner rotation angle of the  wavefront, and $\theta$ is the orbital angle of the particle in the lab frame. From Eq. (\ref{d}) follows that in the ultra relativistic limit $\gamma \longrightarrow \infty$ the wavefront rotates exactly as the velocity vector $\vec{v}$. Generally we denote  successive increments of transverse velocity by $(\Delta v_x)_j$, with $j$ running from 1 to $n$, where the limit for infinitesimally small increments should eventually be taken.
Due to our limitation to the second order, the relativistic corrections to the composition of velocity increments  $(\Delta v_x)_j$ do not appear. We can therefore obtain the total angle of rotation by identifying the velocity change in Eq.(\ref{d}) with the total velocity $v_x$, without looking at differences in clock synchronization between different reference systems. We would then have for the  Wigner rotation angle after the kick $\Phi  \simeq v_x/c$, and the rotation angle of the modulation wave number vector coincides with the angle of rotation of the velocity. We state that in the ultrarelativistic asymptotic the Wigner rotation appears in the first order already, and results directly from the relativity of simultaneity.

In ultrarelativistic asymptotic, the relativistic correction in the composition of velocities  appears already in the second order. In fact, while the relativistic corrections to the composition of velocity increments $(\Delta v_x)_j$ does not appear, the  correction of the longitudinal velocity component appears in this order as $v_z = dz/dt$ $\longrightarrow$ $dz/(\gamma_x dt)$.
In the ultrarelativistic limit we observe  that the resulting Lorentz transformation contains a space (Wigner) rotation which is related with the time shift $t$ $\longrightarrow$  $t - xv_x/c^2$, and a time dilation $t$ $\longrightarrow$ $\gamma_x t$. The former does not enter into the calculation of the velocity increment but appears in the correction to the initial relativistic velocity $\vec{v} = v\vec{e}_z$. The  boost along the $z$-axis imparts the additional velocity $- (v\theta^2/2)\vec{e}_z$ and $\vec{v} = v\vec{e}_z$ $\longrightarrow$ $ v(1 - \theta^2/2)\vec{e}_z$. The Lorentz boost along the $x$-axis gives the time dilation effect. Therefore we finally have $v_z = v(1 - \theta^2/2)$ $\longrightarrow$ $v_z/\gamma_x = v(1 - \theta^2)$.

\subsection{The lab frame view of observations of moving observer}

The laws of physics in any one reference frame should be able to account for all physical phenomena, including the observations made by moving observers. Suppose that we assign absolute space-time coordinates in the lab frame, i.e. a procedure involving slow clock transport is used to assign values of coordinate time. Due to this particular choice of synchronization convention, as thoroughly discussed above, relativistic kinematic effects such as time dilation and length contraction do not exist in the lab frame.

Consider a light clock moving at constant velocity relative to the laboratory observer. Suppose the light clock is moving parallel to the direction of the light pulse. In other words, we assume that the two mirrors move uniformly with the same velocity $v$, in a direction perpendicular to their reflecting surfaces. The light reflection effect caused by the motion of the charges in the mirror plate is all that is required to understand the light clock operation: in other words, from an electrodynamics point of view  we deal with two moving light sources. The first mirror source radiates an electromagnetic wave traveling forward along the positive $z$ direction, while the second one radiates a wave traveling backwards, along the negative $z$ direction. Let us go back to our calculations of the speed of light from the moving source when the clocks in the lab frame are synchronized according to the absolute time convention. When  coordinates are assigned in the lab frame, the laboratory observer can directly measure the one-way speed of light. The result he observes is that the speed of light emitted by the moving source is consistent with the Galilean law of addition of velocities. In particular, when the source is moving with velocity $v$ along the $z$-axis, the velocity of light  in the direction parallel to the $z$-axis, is equal to $c+v$ in the positive, and $-c+v$ in the negative orientations. The principle of relativity assures that no physical (i.e. convention-invariant) observable can depend on the value of $v$. In particular, the principle of relativity requires that the two-way speed of light is equal to $c$ in any given inertial frame. Our next objective is to understand the results of a measurement of  the  two-way speed of light in the case of the  moving  light-clock  described above. Suppose that a traveler, moving with the clock, performs the two-way speed of light measurement. The lab observer sees that the
speed of light relative to the mirrors is $c$ for the wave traveling forward in positive $z$ orientation, and $-c$ for the wave traveling backwards in the negative $z$ orientation. Then, when the measured data is analyzed, the laboratory observer finds that in each case the two-way speed of light is equal to $c$. Due to the Galilean vectorial velocities addition, the laboratory observer will measure the same two-way speed of light, irrespective of the orientation of the mirror clock he is using. The laboratory observer sees from that there is a constant two-way light velocity independently of the speed of the source and its velocity direction. In other words, the measurement of the two-way speed of light is universal and the laboratory observer actually verifies the principle of relativity. In agreement with the principle of relativity, usual Maxwell's equations can be exploited in a moving inertial frame where sources are at rest.

\subsection{Misconception regarding available solutions for covariant electron trajectories}

In general, the covariant equation of motion can be solved only by numerical methods; however, it is always attractive to find instances where exact solutions can be obtained. Let us consider the motion of a particle in a given electromagnetic field. The equation of motion is obtained by specifying  the four-force  as the Lorentz four-force yielding then $dp^{\mu}/d\tau = eF^{\mu\nu}u_{\nu}$, where $p^{\mu} = mu^{\mu}$ is the particle's four-momentum, $\tau$ is its proper time and $u^{\mu}$ its four-velocity. The simplest case, of great practical importance, is that of a uniform electromagnetic field meaning that  $F^{\mu\nu}$ is constant on the whole space-time region of interest. In particular we consider the motion of a particle in a constant homogeneous magnetic field, specified by tensor components $F^{\mu\nu} = B(e^{\mu}_2e^{\nu}_3 - e^{\nu}_2 e^{\mu}_3)$  where $e^{\mu}_2$ and $e^{\mu}_3$ are orthonormal space like basis vectors $e^2_2 = e^2_3 = - 1$, $e_2\cdot e_3 = 0$. In the lab frame of reference where $e^{\mu}_0$ is taken as the time axis, and   $e^{\mu}_2$ and $e^{\mu}_3$ are space vectors the field is indeed purely magnetic, of magnitude $B$ and parallel to the $e_1$ axis. Let us set the initial four-velocity $u^{\mu}(0) = \gamma c e^{\mu}_0 + \gamma v e^{\mu}_2$, where $v$ is the initial particle's velocity relative to the lab observer along the axis $e_2$ at the instant $\tau = 0$, and $\gamma = 1/\sqrt{1-v^2/c^2}$. The components of the equation of motion are then $du^{(0)}/d\tau = 0$, $du^{(1)}/d\tau = 0$, $du^{(2)}/d\tau = - eBu^{(3)}/(mc)$, $du^{(3)}/d\tau =  eBu^{(2)}/(mc)$. We seek for the initial value solution to these equations as  done in the existing literature (see e.g. \cite{RM,GF,CT}). A distinctive feature of the initial value problem in relativistic mechanics, is that the dynamics is always constrained. In fact, the evolution of the particle is subject to $dp^{\mu}/d\tau = eF^{\mu\nu}u_{\nu}$, but also to the constraint $u^2 = c^2$. However, such a condition can be weakened requiring its validity at certain values of $\tau$ only, let us say initially, at $\tau = 0$. To prove this,
we make a scalar product of the equation of motion by $u_{\mu}$. Using the fact that  $F^{\mu\nu}$ is antisymmetric (i.e. $F^{\mu\nu} = - F^{\nu\mu}$), we find $u_{\mu}d u^{\mu}/d\tau = eF^{\mu\nu}u_{\mu}u_{\nu} = 0$. Thus, for quantity $Y = (u^2 - c^2)$ we find $dY/d\tau = 0$. Therefore, if $Y(\tau)$ vanishes initially, i.e. $Y(0) = 0$, then $Y(\tau) = 0$ at any $\tau$. In other words, the differential Lorentz-force equation implies the constraint $u^2 = c^2$ once this is satisfied initially \cite{RCM}. Integrating with respect to the proper time we have  $u^{\mu}(\tau) = \gamma e^{\mu}_0 + \gamma v[e^{\mu}_2\cos(\omega\tau)
+ e^{\mu}_3\sin(\omega\tau)]$ where $\omega = eB/(mc)$. We see that $\gamma$ is constant with time, meaning that the energy of a charged particle moving in a constant magnetic field is constant. After two successive integrations we have $X^{\mu}(\tau) = X^{\mu}(0) + \gamma c\tau e^{\mu}_0 + R[e^{\mu}_2 \sin(\omega\tau) - e^{\mu}_3\cos(\omega\tau)]$ where $R = \gamma v/\omega$.  This enables us to find the time dependence $[0,X^{(2)}(t), X^{(3)}(t)]$ of the particle's position since $t/\gamma = \tau$.  From this solution of the equation of motion we conclude that the motion of a charged particle in a constant magnetic field is a uniform circular motion.

One could expect that the particle's trajectory  $[0, X^{(2)}(t), X^{(3)}(t)]$ in the lab frame, following from the previous reasoning, should be identified with $\vec{x}_{cov}(t)$. However, paradoxical result may be obtained by doing so. In particular, the trajectory $[0,X^{(2)}(t), X^{(3)}(t)]$  does not include relativistic kinematics effects. The point is that $[0,X^{(2)}(t), X^{(3)}(t)]$ cannot be identified with $\vec{x}_{cov}(t)$ even if, at first glance, it appears to be derived following our covariant prescription. In fact, the calculation carried out above shows that  $t/\gamma = \tau$  and one can see the connection between this dependence  and the absolute simultaneity convention. Here we have a situation where the temporal coincidence of two events has the absolute character:  $\Delta \tau = 0$ implies $\Delta t = 0$.

The reason that time differs from space is due to the fact that the particle trajectory was found by integration from initial conditions. In other words, the study of relativistic particle motion was identified with a well-defined initial value problem. This approach  is based on the hidden assumption that the type of clock synchronization, which provides the time coordinate $t$ in the lab frame, is based on the use of the absolute simultaneity convention.

The definition of the time coordinate is important to provide a starting point for the algorithm  of reconstruction of the particle trajectory. Based on the structure of the four components of the equation of motion,  we can  arrive to another mathematically identical formulation of the initial value problem. The fact that the evolution of the particle in the lab frame is subject to a constraint has already been mentioned. This means that the mathematical form of the dynamics law includes only three independent equations of motion. It is easy to see from the initial set of four equations that it is possible to present the time component simply as the relation $d\tau = dt/\gamma$. Actually, it is just a simple parametrization that yields the corrected Newton's equation Eq.(\ref{N}) as another equivalent form of these four equations in terms of absolute time $t$ instead of proper time of the particle. This approach to integrating dynamics equations from the initial conditions relies on the use of three independent spatial coordinates and velocities without constraint.

Here the fundamentally important point to consider is that  the clock synchronization procedure within the lab frame, which provides the time coordinates $t$ in the corrected Newton's equation, is based on the use the absolute time convention.

Let us now consider a relativistic particle, accelerating in the lab frame, and let us analyze its evolution within the Lorentz coordinate system. The  permanent rest frame of the particle is obviously not inertial and any transformation of observations in the lab frame, back to the rest frame, cannot be made by means of Lorentz transformations. To get around that difficulty one introduces an infinite sequence of co-moving frames. At each instant, the rest frame is a Lorentz frame  centered on the particle and moving with it. As the particle velocity changes to its new value at an infinitesimally later instant,  a new Lorentz frame centered on the particle and moving with it at the new velocity is used to observing the particle. All reference frames are assumed to be orthogonal. This ensemble of comoving coordinate systems or tetrads can be constructed by choosing, for each value of $\tau$ along the world line $\sigma$ of the particle, an inertial  system  whose origin coincides with $\sigma(\tau)$ and whose $x'_0$-axis is tangent to $\sigma$ at $\sigma(\tau)$. The zeroth basis vector $e'_0$ is therefore directed as the 4-velocity $u$.
In the tetrad basis $e'_i(\tau)$, the particle has four velocity $u = (c, 0, 0, 0)$ and four acceleration $a = (0,a_1,a_2,a_3)$.
The basis vectors of the tetrad  $e'_0(\tau), e'_1(\tau), e'_2(\tau), e'_3(\tau)$ at any proper time $\tau$ are then related to the basis vectors $e_0, e_1, e_2, e_3$ of some given inertial lab frame by a Lorentz transformation $e'_\mu(\tau) = \Lambda^\nu_\mu(\tau)e_\nu$. Therefore, the basis vectors at two successive instants must also be related to each other by a Lorentz transformation.

In the lab frame one thus has a coordinate representation of the world-line as  $\sigma(\tau) = (t(\tau), x_1(\tau), x_2(\tau), x_3(\tau))$. The covariant particle trajectory $\vec{x}_{cov}(t)$ is calculated by projecting world line  to the lab frame basis  and using the lab time $t$ as a parameter for the trajectory curve.
In this paper we claimed many times that there is a difference between the non-covariant particle trajectory $\vec{x}(t)$, calculated by solving the corrected  Newton's equations and the covariant particle trajectory $\vec{x}_{cov}(t)$, calculated by projecting the world line onto the  lab frame Lorentz basis. There is a fundamental reason for this difference. The trajectory $\vec{x}_{cov}(t)$ is viewed from the lab frame as the result of Lorentz transformations $\Lambda^\nu_\mu(\tau)$ that depend on the proper time. Therefore, the composition law that follows from the group properties of the Lorentz transformations  is used to express the conditions of co-moving sequence of frames tracking a particle. In contrast to this, $\vec{x}(t)$ follows from solving the corrected Newton's equations and does not include Lorentz transformation composition law.

Let us discuss an explicit example of the difference between $\vec{x}(t)$ and $\vec{x}_{cov}(t)$ trajectories. As is known, the composition of non-collinear Lorentz boosts does not result in a different boost but in a Lorentz transformation involving a boost and a spatial rotation, the Wigner rotation. Suppose that our  particle moves along an arbitrary accelerated world line. As just discussed, the basis vectors of the tetrad defining the instantaneously co-moving frames is related to the basis vectors of the lab frame  by a  Lorentz transformation depending on the proper time $e'_\mu(\tau) = \Lambda^\nu_\mu(\tau)e_\nu$. The most general Lorentz transformation  $\Lambda^\nu_\mu(\tau)$  can be  uniquely separated into a pure Lorentz boost followed by spatial rotation. As seen from the lab frame, space vectors of the tetrad (those with indexes $\mu = 1,2,3$) rotate relative to  the Cartesian axes of the lab frame. The Wigner rotation does not occur due to the action of some forces and has pure relativistic kinematics origin.

Let us try out our algorithm for reconstructing $\vec{x}_{cov}(t)$ on some example, to see how it works. Suppose that a particle is moving with velocity $\vec{v}$ in the lab frame for an instant. At this $\tau = 0$ instant one picks a Lorentz coordinates frame. Let the $S$ be a lab frame of reference, $S'$ a comoving with velocity $\vec{v}$ relative to $S$. It is assumed that $S'$ is connected to $S$ by a Lorentz transformation: $X'_{\mu} = \Lambda^\nu_\mu(0)X_{\nu}$. Then, in the time interval $\Delta t$, the particle velocity changes of a small value $\Delta\vec{v}$, say, along the $x$-axis. At this first step Eq. (\ref{DE1}) allows one to express the velocity increment $\Delta \vec{v}$ through the time interval $\Delta t = \gamma\Delta \tau$ in the initial Lorentz coordinate system. In order to keep Lorentz coordinates in the lab frame, however, one additionally needs to perform the Lorentz boost $L(\Delta\vec{v})$. The correct transformation, for $\Delta v/c$ so small that $(\Delta v/c)^2$ is neglected, is seen to be given by $x \longrightarrow  x - \Delta v t$ and $t  \longrightarrow t - x\Delta v/c^2$. In the subsequent  time interval, the  equation of motion  Eq.(\ref{DE1}) is valid in the Lorentz lab frame whose basis is related to the comoving frame basis by the  Lorentz transformation $e'_\mu(\Delta\tau) = \Lambda^\nu_\mu(0)L(\Delta\vec{v})^\alpha_\nu e_\alpha$. To obtain a transformation valid for a finite proper time interval $\tau$ we must consider $n$ successive transformations $\Lambda(0)L(\Delta\vec{v_1})L(\Delta\vec{v_2}) ... L(\Delta\vec{v_n})$ and then take the limit $n \longrightarrow \infty$, $n\Delta \tau  \longrightarrow \tau$.  As just seen, the trajectory  $\vec{x}_{cov}(t)$ is viewed from the Lorentz lab frame as a result of successive  Lorentz transformations.  In Lorentz coordinates, the lab time $t$ in the equation of motion Eq.(\ref{DE1}) cannot be independent from the space variables.  This is because resynchronization of distant clocks in the process of particle acceleration leads to a mixture of positions and time.

\section{Conclusions}

A non-covariant (3+1) approach to relativistic particle dynamics has been used in particle tracking calculations for about  seventy years. However, the type of  clock synchronization which provides the time coordinate $t$ in the corrected Newton's equation  has never been discussed in literature. We claim, and this claim is quite central for our reasoning, that in conventional particle tracking in accelerator and plasma physics the description of dynamical evolution in the lab frame is based on the use of the absolute time convention.

In the theory of relativity this choice may seem quite unusual, but it is usually the most convenient one in relativistic engineering.
In non covariant particle tracking,  time differs from space and  a particle trajectory can be seen from the lab frame as the result of successive Galilean boosts that track the accelerated motion. The usual Galileo (vectorial) rule for addition of velocities is used to fix Galileo boosts tracking a particular particle along its motion.

The usual approach to relativistic charged particle dynamics in accelerator and plasma physics relies on physics concepts that do not require the introduction of a four-dimensional Minkowski space.
The conventional particle tracking in accelerator and plasma physics actually parallels non-relativistic ideas introducing,  as only modification, the relativistic mass.
The dynamics of charged particles is described by the conventional  3-Lorentz force since the evolution parameter is, like in non-relativistic dynamics, the absolute time $t$.  Actually, from an operational point of view, it is assumed that the simplest self-evident procedure of slow clock transport is used to assign numerical values to the time coordinate in the lab frame. Therefore, in this approach we have  no mixture of positions and time. There is a reason to prefer non-covariant way within the framework of dynamics only. The non-covariant approach to particle dynamics relies on the use of three independent coordinates and velocities without constraints (such as $u^2 = c^2$ in the manifestly covariant approach). This approach should be used in the study of relativistic particle motion in a prescribed force field, since  it is   a well-defined initial value problem \cite{RM}. This (3+1) dimensional non-covariant particle tracking method  is simple, self-evident,  and adequate to the laboratory reality. In fact, our experimental apparatuses function in three-dimensional space and one-dimensional time. However, we are better off using covariant trajectories when we want to solve the electrodynamics problem based on Maxwell's equations in their usual form. As we have seen, in fact, the use of non-covariant trajectories also implies the use of much more complicated electromagnetic field equations.

For the first time we showed a difference between conventional and covariant particle tracking results in the lab frame. This essential point has never received attention in the physical community. Only the solution of the dynamics equations in covariant form gives the correct coupling between the usual Maxwell's equations and particle trajectories in the lab frame. We conclude that previous theoretical and experimental results in accelerator and plasma physics should be reexamined in the light of the pointed difference between conventional and covariant particle tracking. In particular, a correction of the conventional synchrotron radiation theory is required. One can see that the difference between conventional particle trajectory and covariant particle trajectory seems to have been entirely overlooked by many physicists including J. Schwinger \cite{SCH}, who developed results of synchrotron radiation theory by mistakenly using the usual Maxwell's equations and $\vec{x}(t)$, instead of $\vec{x}_{cov}(t)$.

We presented a study of an experimental  setup for illustrating the difference between conventional and covariant trajectories. We solved the dynamics problem of the motion of a relativistic electron in the prescribed force field of a weak dipole magnet by working only up to the order $(\gamma\theta)^2$. This approximation is of particular theoretical interest because it is relatively simple and at the same time forms the basis for understanding  relativistic kinematics effects such as Wigner rotation and time dilation. This study  has also practical applications. The second order approximation used to investigate the kicker setup in this paper could be used in a large variety of practical problems in XFEL engineering.

\end{document}